\documentclass[11pt,a4paper]{article}
\usepackage[T1]{fontenc}
\usepackage[utf8]{inputenc}
\usepackage[english]{babel}
\usepackage[numbers,sort&compress]{natbib}
\usepackage{amsmath}
\usepackage{amssymb}
\usepackage{amsthm}
\usepackage{times}
\usepackage{a4}
\usepackage[hang,stable]{footmisc}

\usepackage{xspace}
\usepackage{graphicx}
\usepackage{color}
\usepackage{bbm}
\usepackage{slashed}
\usepackage{units}

\usepackage{caption}
\newcommand{\sne}{Schr\"o\-din\-ger-New\-ton equation\xspace}
\newcommand{\kg}{Klein-Gordon equation\xspace}
\newcommand{\e}{\mathrm{e}}

\newcommand{\abs}[1]{\vert #1 \vert}

\newcommand{\1}{\mathbbm{1}}
\newcommand{\D}{\mathrm{d}}
\newcommand{\bleq}{\mathrel{\phantom{=}}}
\newcommand{\order}[1]{\mathcal{O}(#1)}
\newcommand{\grad}{\vec{\nabla}}

\newcommand{\akomm}[2]{\left\{#1,#2\right\}}
\newcommand{\tr}[1]{\mathrm{Tr}\left(#1\right)}

\newcommand{\tmatrix}[4]{\left(\begin{array}{cc}#1&#2\\#3&#4\end{array}\right)}
\begin{document}
\ifx\href\undefined\else\hypersetup{linktocpage=true}\fi

\title{The Schr\"odinger-Newton equation as non-relativistic limit of self-gravitating
       Klein-Gordon and Dirac fields}
\author{Domenico Giulini and Andr\'e Gro{\ss}ardt       \\
        Center of Applied Space Technology and Microgravity\\
        University of Bremen             \\
        Am Fallturm 1                    \\
        D-28359 Bremen, Germany          \\
        and                              \\
        Institute for Theoretical Physics\\
        Leibniz University Hannover      \\
        Appelstrasse 2                   \\
        D-30167 Hannover, Germany}

\date{}

\maketitle

\begin{abstract}
\noindent In this paper we show that the \sne for spherically symmetric
gravitational fields can be derived in a WKB-like expansion 
in $1/c$ from the Einstein-Klein-Gordon and Einstein-Dirac 
system.
\end{abstract}

\section{Introduction}
This work is the sequel to a recent study~\cite{Giulini.Grossardt:2011} 
in which we analysed the qualitative and quantitative behaviour 
of Gaussian wave-packets moving according to the time-dependent \sne:
\begin{equation}
\label{eq:SchroedingerNewton}
\mathrm{i} \hbar \, \partial_t \psi(t,\vec x) =
  \left(-\frac{\hbar^2}{2m} \Delta - G m^2
  \int\frac{\vert\psi(t,\vec y)\vert^2}{\Vert\vec x-\vec y\Vert}\,\D^3 y
  \right) \psi(t,\vec x) \,.
\end{equation}
In this paper we shall be concerned with the question of 
how \eqref{eq:SchroedingerNewton} can be understood as a 
consequence of known principles and equations.   

We recall that originally  \eqref{eq:SchroedingerNewton} 
was suggested in \cite{Diosi:1984,Penrose:1998} as a model 
for the gravitational localisation of macro objects. 
This was considered to be an essentially new mechanism that 
also involves hitherto unknown physics. In that sense \eqref{eq:SchroedingerNewton} was taken by many as a 
hypothesis to be put to test by experiment. It is then natural 
to ask whether it predicts any observable consequences, say in 
molecular interferometry, that have any chance of being 
detected in the foreseeable future. It may also be envisaged that 
it might give interesting hints concerning the interface 
between quantum physics and gravitation, and that it might 
even shed some light on the intricate question concerning the 
necessity of quantum gravity~\cite{Carlip:2008,Salzman.Carlip:2006}.

Our results in \cite{Giulini.Grossardt:2011} showed that on 
the basis of the \sne inhibitions of the dispersion of wave 
packets due to their own gravitational field start to become 
significant at mass scales around $\unit[10^{10}]{u}$ for width of 
$\unit[500]{nm}$. This is more than 6 orders of magnitude above the 
current masses in molecular interferometry~%
\cite{Arndt.Hornberger.Zeilinger:2005,HornbergerEtAl:2012} 
but only 3  orders beyond the masses already envisaged possible 
in future experiments.

This paper is complementary to these attempts and devoted to 
a better \emph{foundational} understanding of the \sne, which we 
shall view as a general model for gravitational self-interaction 
of matter waves. Accordingly, we shall discuss its derivability 
from known principles, equations, and approximation schemes. 
Through that procedure we also hope to gain insight into the 
presently justifiable range of applicability of 
\eqref{eq:SchroedingerNewton}, which has been a matter of some 
debate over recent years. Needless to say that our approach 
here does of course not exclude the logical  possibility that \eqref{eq:SchroedingerNewton} may capture new physics, as 
originally anticipated. But even then it would be good to know 
what aspects of it just represent old physics. 

The question we pose closely relates to that of how classical 
gravitational fields couple to ``quantum matter'', i.\,e. matter 
that is described by the Schr\"odinger equation. Note that here 
``gravitational fields'' include those sourced by the quantum 
matter itself. For classical fields the coupling prescription 
that derives from the Equivalence Principle is the 
\emph{minimal coupling scheme}, which proceeds in three steps:
\begin{enumerate} 
\item
Formulate the theory of the field you wish to couple to 
gravity in a way that satisfies the principles of Special 
Relativity. In particular: write the field equations 
in a Poincar\'e invariant fashion. 
\item
Replace the Minkowski metric by a general Lorentzian 
metric $g$ and the Levi-Civita covariant derivative w.\,r.\,t.
the Minkowski metric (which is just the partial derivative in 
global inertial coordinates in Minkowski space) by the
Levi-Civita covariant derivative for the general metric $g$.
\item
Calculate the energy-momentum tensor $T$ for the matter field
from the result of step\,2 through variation of the action 
with respect to $g$ and impose on $g$ Einstein's gravitational
field equation with a right-hand side (energy-momentum tensor) 
including $T$ (and possible external sources). 
\end{enumerate}
There are some ambiguities in step~2 which are well known
to result in differences in the couplings to the curvature 
tensor. These are not essential for our concern here 
and will be ignored. In contrast, far more serious for us 
are the difficulties posed by step\,1. It requires the 
formulation of special-relativistic quantum mechanics, the 
``non-relativistic limit''  ($1/c\rightarrow 0$) of which 
is Schr\"odinger's theory.%
\footnote{In this paper we follow standard practise in  
speaking of ``non-relativistic'' approximations or limits, 
even though this is unfortunate terminology since it really 
means to approximate Poincar\'e symmetry by Galilei 
symmetry, both implementing the relativity principle.} 
However, if ``special-relativistic quantum mechanics'' is 
interpreted as ``special-relativistic one-particle quantum 
mechanics'' this is impossible as such a theory is known 
not to exist (particle creation and annihilation). This 
seems to force us to interpret ``special-relativistic quantum 
mechanics'' as ``Relativistic Quantum Field Theory''. In that 
case step~2 would presumably be achieved for \emph{external} 
gravitational fields by considering Quantum Field Theory 
in curved space, but the gravitational self interaction 
could certainly not be captured in that way. 

We conclude that, strictly speaking, there is no obvious way to 
deduce from the standard formulation of the Equivalence Principle 
the general coupling of a classical gravitational field to matter 
described by Schr\"odinger's equation. For a spatially homogeneous gravitational field $\vec g(t)$ with arbitrary time dependence 
one could deduce the standard coupling, which consists of an 
addition of $V(\vec x,t)=m\vec g(t)\cdot\vec x$ to the potential 
in the Hamiltonian, from the standard requirement of equivalence 
with rigidly accelerated frames (see, e.\,g., \cite{Giulini:2012x}).
This has recently also be shown to work if one starts from 
the Klein-Gordon equation written in co-moving coordinates of
a rigidly but time dependent accelerating frame and taking 
the appropriate $1/c\rightarrow 0$ limit~\cite{Padmanabhan.Padmanabhan:2011}.
But, to stress it once more, the general case does not seem to 
be deducible by standard procedures.

What we will do is to follow the example just mentioned and 
\emph{formally} regard the Schr\"odinger equation as the 
non-relativistic limit of the Klein-Gordon or Dirac equation. 
To be sure, this is just what is often done in textbooks on 
Quantum Mechanics, and there is indeed nothing to be upset 
about as long as this is understood strictly as approximation 
scheme for differential equations of complex-valued fields 
on spacetime. However, what remains to be clarified is the 
interpretation of both fields, and their relation in the 
non-relativistic limit. This, we feel, is an important issue 
that should be given due thought before rushing into suggestions
to look for possible experimental signatures of \sne. In the 
present paper, however,  we concentrate on the formal problem 
and will only shortly come back to the interpretational issues 
at the end. 

\section{The non-relativistic approximation scheme}
We systematise the notion of ``non-relativistic approximation''
by following the WKB-type procedure of Kiefer \& Singh 
\cite{Kiefer:1991}. Recall that the traditional WKB approximation 
is a scheme to obtain the semi-classical limit of a quantum 
theory by means of a formal expansion in terms of a dimensionful 
parameter $\hbar$, which is Planck's constant $h$ divided by 
$2\pi$. It starts by inserting the ansatz
\begin{equation}
 \psi(\vec{x},t) \sim \exp\left(\frac{\mathrm{i}}{\hbar} S(\vec{x},t)\right)
\end{equation}
for the wave function into a given linear partial differential 
equation, e.\,g. the Schr\"odinger or Klein-Gordon equation, 
and subsequently expanding the exponent $S$ in terms of 
$\hbar$ \cite{Wentzel:1926}. The equation is then required to be 
satisfied at each order in $\hbar$. At each order one obtains a 
certain truncation of the original theory, which makes mathematical 
sense and may or may not lead to good approximations of the latter.
In this sense $\hbar$ should be viewed as a deformation parameter 
of the theory rather than an approximation parameter. In the latter 
case we have to worry about convergence of this expansion, where 
the degree of smallness assigned to $\hbar$ will depend on the 
context. The formal expansion, however, makes sense independent 
of any context. 

It was shown by Pauli \cite{Pauli:1932} how this scheme can be adopted
to wave functions with multiple components, as for the Dirac equation.
Pauli derived as the semi-classical limit of the Dirac equation a set of
equations that he could not solve in general. This was completed by 
\citet{Rubinow:1963} and later by \citet{Rafanelli:1964}, as well as
\citet{Pardy:1973}, to yield, e.\,g.,  the classical 
Bargmann-Michel-Telegdi (BMT) equation~\cite{Bargmann:1959} at the 
appropriate order in $\hbar$. An alternative method to obtain the semiclassical limit of the Dirac equation using matrix-valued Wigner functions, which also leads to the BMT equation, was more recently 
developed by \citet{Spohn:2000}.

Similarly to the traditional WKB-method, the concept of a 
non-relativistic limit of a relativistic field theory 
can be understood as appropriate order in an expansion of 
the dimensionful parameter $1/c$. Again $1/c$ should be viewed 
as deformation parameter linking different theories. In particular 
it allows to contract the Poincar\'e symmetric theory to a 
Galilei symmetric theory as $1/c\rightarrow 0$. This results 
in a WKB-like scheme applied to the parameter $1/c$.
\citet{Kiefer:1991} showed that this can be used to derive 
the Schr\"odinger equation from the Klein-Gordon equation, 
and then generalised this method to derive quantum gravity 
corrections from the functional Wheeler-de\,Witt equation.

We argue in some detail that this method provides a universal 
scheme in which $\hbar$ as well as $1/c$ act as deformation 
parameters, and show that the \sne occurs as a $1/c\rightarrow 0$ 
limit of the Klein-Gordon and Dirac field coupled to Einstein 
gravity. This goes beyond a former analysis of \citet{Oliveira:1962}
which uses the traditional method of Foldy-Wouthuysen \cite{Foldy:1950}
but does not recouple the Dirac field into Einstein's equations.

Following the strategy just outlined, we can make the ansatz
\begin{equation}
\label{eqn:ansatz-fields}
 \psi(\vec{x},t) = \exp\left(\frac{\mathrm{i} c^2}{\hbar} S(\vec{x},t)\right)
 \sum_{n=0}^\infty \left(\frac{\sqrt{\hbar}}{c}\right)^n a_n(\vec{x},t),
\end{equation}
where $\psi$ can be a scalar, vector or spinor field and 
so are the $a_n$, but $S$ always is a scalar function. 
The semi-classical and non-relativistic limits can then 
be derived by inserting this ansatz into the field equation 
considered and sorting by powers of either $\hbar$ or $1/c$.

We will perform both the semi-classical and the non-relativistic 
limit for the Klein-Gordon equation with an electromagnetic 
field in paragraph \ref{par:kg}. Coupling the Klein-Gordon 
equation to general relativity we show that the \sne can be 
derived as the non-relativistic limit of the coupled 
Einstein-Klein-Gordon system.

In paragraph \ref{par:dirac} we will repeat this analysis 
for the Dirac equation with an electromagnetic field. We 
obtain the known results, namely the BMT equation as the 
semi-classical and the Pauli equation as the non-relativistic 
limit, respectively. When coupled to Einstein's equations,
the non-relativistic limit of the Dirac equation yields 
the \sne, too.

Throughout our signature convention for the metric will 
be ``mostly plus'', i.\,e. $(-,+,+,+)$.

\section{Klein-Gordon fields} \label{par:kg}
The free Klein-Gordon equation reads
\begin{equation}
 \label{eqn:free-KG}
 \left( \Box - \frac{m^2 c^2}{\hbar^2} \right) \psi = 0, \hspace{2cm} \Box
  = -\frac{1}{c^2} \partial_t^2 + \Delta,
\end{equation}
where $\psi$ is a scalar field. Introducing an electromagnetic field with 
electric potential $\phi$ and vector potential $\vec{A}$ by replacing
\begin{subequations}%
\begin{alignat}{2}
 \partial_t &\rightarrow \partial_t &&+ \frac{\mathrm{i} e}{\hbar} \phi(\vec{x},t) \\
 \partial_k &\rightarrow \partial_k &&- \frac{\mathrm{i} e}{\hbar} A_k(\vec{x},t) \\
\label{eqn:Box-Operator}
 \Box &\rightarrow \Box &&+ \frac{e^2}{\hbar^2 c^2}
  \left(\phi^2 - c^2 A^2\right)
  - \frac{2 \mathrm{i} e}{\hbar c^2} \left(\phi \partial_t + c^2 \vec{A} \cdot 
  \vec{\nabla}\right) \nonumber\\
 &&&- \frac{\mathrm{i} e}{\hbar c^2} \left(\dot{\phi} + c^2 \grad \cdot \vec{A}\right),
\end{alignat}%
\end{subequations}
the Klein-Gordon equation takes the form
\begin{multline}
\label{eqn:kg}
 \Big( \partial_t^2 - c^2 \Delta + \frac{2 \mathrm{i} e}{\hbar} \left( \phi \partial_t
  + c^2 \vec{A} \cdot \vec{\nabla} \right) - \frac{e^2}{\hbar^2} \left( \phi^2
  - c^2 A^2 \right) + \frac{m^2 c^4}{\hbar^2} \\+ \frac{\mathrm{i} e}{\hbar} \left( 
  \dot{\phi} + c^2 \vec{\nabla} \cdot \vec{A} \right) \Big) \psi = 0.
\end{multline}
Note that the last lines of equations \eqref{eqn:Box-Operator} and
\eqref{eqn:kg} could be cancelled using the Lorenz gauge
$\dot{\phi}+c^2 \grad \cdot \vec{A}=0$, but we do not want to 
fix a gauge at this stage because it would be $c$-dependent.

We now make use of the ansatz \eqref{eqn:ansatz-fields} and 
calculate the first and second order temporal and spatial 
derivatives for the field $\psi$. They are
\begin{subequations}%
\label{eqn:ansatz-derivatives}%
\begin{alignat}{2}
 \partial_t \psi &= \e^{\mathrm{i} c^2 S / \hbar}\, \frac{c^2}{\hbar}
  \sum_{n=0}^\infty \left(\frac{\sqrt{\hbar}}{c}\right)^n
  & \Big( &\mathrm{i} \dot{S} a_n + \dot{a}_{n-2} \Big) \\
 \grad \psi &= \e^{\mathrm{i} c^2 S / \hbar}\, \frac{c^2}{\hbar} \sum_{n=0}^\infty 
  \left(\frac{\sqrt{\hbar}}{c}\right)^n & \Big( &\mathrm{i} (\grad S) a_n
  + \grad a_{n-2} \Big) \\
 \partial_t^2 \psi &= \e^{\mathrm{i} c^2 S / \hbar}\, \frac{c^4}{\hbar^2}
  \sum_{n=0}^\infty \left(\frac{\sqrt{\hbar}}{c}\right)^n 
  &\Big( &- \dot{S}^2 a_n + 2 \mathrm{i} \dot{S} \dot{a}_{n-2} 
  + \mathrm{i} \ddot{S} a_{n-2} + \ddot{a}_{n-4} \Big) \\
 \Delta \psi &= \e^{\mathrm{i} c^2 S / \hbar}\, \frac{c^4}{\hbar^2} \sum_{n=0}^\infty 
  \left(\frac{\sqrt{\hbar}}{c}\right)^n &\Big( &- (\grad{S})^2 a_n
  + 2 \mathrm{i} (\grad{S}) \cdot \grad a_{n-2} \nonumber\\
 &&& + \mathrm{i} (\Delta S) a_{n-2} + \Delta a_{n-4} \Big),
\end{alignat}%
\end{subequations}
where we denote the time derivative $\partial_t$ by a 
dot and define $a_n \equiv 0$ for all $n<0$.
Inserting this ansatz into the \kg \eqref{eqn:kg} yields
\begin{align}
\label{eqn:ansatz-kg}
 0 &= \exp\left(\frac{\mathrm{i} c^2}{\hbar} S\right) \frac{c^4}{\hbar^2}
  \sum_{n=0}^\infty \left(\frac{\sqrt{\hbar}}{c}\right)^n \Big[ -\dot{S}^2 a_n
  + 2 \mathrm{i} \dot{S} \dot{a}_{n-2} + \mathrm{i} \ddot{S} a_{n-2} + \ddot{a}_{n-4}
  \nonumber\\
 &\bleq + c^2 (\grad S)^2 a_n - 2 \mathrm{i} c^2 (\grad S) \cdot \grad a_{n-2}
  - \mathrm{i} c^2 (\Delta S) a_{n-2} - c^2 \Delta a_{n-4}  \nonumber\\
 &\bleq - \frac{2 e \phi}{\hbar} \dot{S} a_{n-2} + \frac{2 \mathrm{i} e \phi}{\hbar} 
  \dot{a}_{n-4} - 2 e (\vec{A} \cdot \grad S) a_n
  + 2 \mathrm{i} e \vec{A} \cdot \grad a_{n-2} \nonumber\\
 &\bleq - \frac{e^2 \phi^2}{\hbar^2} a_{n-4} + \frac{e^2 A^2}{\hbar} a_{n-2}
  + m^2 a_n + \frac{\mathrm{i} e}{\hbar} \dot{\phi} a_{n-4}
  + \mathrm{i} e (\grad \cdot \vec{A}) a_{n-2} \Big].
\end{align}

\subsection{The semi-classical limit}
We first rewrite \eqref{eqn:ansatz-kg} by eliminating 
the $\hbar$ dependence inside the square brackets through 
appropriately shifting the summation index of the terms 
containing $\hbar$:
\begin{align}
 0 &= \exp\left(\frac{\mathrm{i} c^2}{\hbar} S\right) \frac{c^4}{\hbar^2}
  \sum_{n=0}^\infty \left(\frac{\sqrt{\hbar}}{c}\right)^n \Big[ 
  -\dot{S}^2 a_n + 2 \mathrm{i} \dot{S} \dot{a}_{n-2} + \mathrm{i} \ddot{S} a_{n-2} 
  + \ddot{a}_{n-4}  \nonumber\\
 &\bleq + c^2 (\grad S)^2 a_n - 2 \mathrm{i} c^2 (\grad S) \cdot \grad a_{n-2} 
  - \mathrm{i} c^2 (\Delta S) a_{n-2} - c^2 \Delta a_{n-4}  \nonumber\\
 &\bleq - \frac{2 e \phi}{c^2} \dot{S} a_{n} + \frac{2 \mathrm{i} e \phi}{c^2}
  \dot{a}_{n-2} - 2 e (\vec{A} \cdot \grad S) a_n + 2 \mathrm{i} e \vec{A} 
  \cdot \grad a_{n-2} \nonumber\\
 &\bleq - \frac{e^2 \phi^2}{c^4} a_{n} + \frac{e^2 A^2}{c^2} a_{n} 
  + m^2 a_n + \frac{\mathrm{i} e}{c^2} \dot{\phi} a_{n-2} 
  + \mathrm{i} e (\grad \cdot \vec{A}) a_{n-2} \Big]\,.
\end{align}
Sorting by powers of $n$ we obtain equations
\begin{align}
 0 &= \left(m^2 - \dot{S}^2 + c^2 (\grad S)^2 - \frac{2 e \phi}{c^2} \dot{S} 
  - 2 e (\vec{A} \cdot \grad S) - \frac{e^2 \phi^2}{c^4} 
  + \frac{e^2 A^2}{c^2}\right) a_n \nonumber\\
 &\bleq + \mathrm{i} \left(\ddot{S} - c^2 (\Delta S) + \frac{e}{c^2} \left( \dot{\phi} 
  + c^2 \grad \cdot \vec{A} \right)\right) a_{n-2} + 2 \mathrm{i} \left( \dot{S} 
  + \frac{e \phi}{c^2} \right) \dot{a}_{n-2} \nonumber\\
 &\bleq - 2 \mathrm{i} c^2 \left((\grad S) - \frac{e \vec{A}}{c^2}\right) 
  \cdot \grad a_{n-2} + \ddot{a}_{n-4} - c^2 \Delta a_{n-4}\,,
\end{align}
one for each $n$. For $n=0$ this yields
\begin{equation}
 0 = (m c^2)^2 - \left(c^2 \dot{S} + e \phi\right)^2 
  + c^2 \left(c^2 \grad S - e \vec{A}\right)^2\,,
\end{equation}
which is the Hamilton-Jacobi equation for a relativistic 
particle. The equations can be simplified further if 
we introduce the four vector $\pi_\mu$ with 
$\pi_0 = -c \dot{S} - e \phi / c$ and 
$\pi_k = -c^2 \partial_k S + e A_k$. We obtain
\begin{equation}
 0 = m^2 c^2 + \pi_\mu \pi^\mu.
\end{equation}

At order $n=2$, now making use of the Lorenz gauge, 
we get
\begin{equation}
 0 = \left(c^2 \ddot{S}-c^4 \Delta S\right) a_0 + 2 \left(c^2 \dot{S} 
  + e \phi\right) \dot{a}_0 - 2 c^2 \left(c^2 \grad S 
  - e \vec{A}\right) \cdot \grad a_0\,,
\end{equation}
which (with $\partial_0 = \partial_t / c$) can be written as
\begin{equation}
 0 = (\partial_\mu \pi^\mu) a_0 + 2 \pi_\mu \partial^\mu a_0.
\end{equation}

\subsection{The non-relativistic limit}
Again we rewrite equation \eqref{eqn:ansatz-kg}, but this 
time we eliminate the $c$ dependence inside the square
brackets through appropriately shifting the summation index 
of the terms containing $c$. In order not to list terms
with index larger than $n$, we make an overall shift 
$n\rightarrow (n-2)$ and compensate for this by an 
overall multiplication with $c^2/\hbar$:
\begin{align}
 0 &= \exp\left(\frac{\mathrm{i} c^2}{\hbar} S\right) \frac{c^6}{\hbar^3} 
  \sum_{n=0}^\infty \left(\frac{\sqrt{\hbar}}{c}\right)^n 
  \Big[ -\dot{S}^2 a_{n-2} + 2 \mathrm{i} \dot{S} \dot{a}_{n-4} 
  + \mathrm{i} \ddot{S} a_{n-4} + \ddot{a}_{n-6}  \nonumber\\
 &\bleq + \hbar (\grad S)^2 a_n - 2 \mathrm{i} \hbar (\grad S) \cdot \grad a_{n-2} 
  - \mathrm{i} \hbar (\Delta S) a_{n-2} - \hbar \Delta a_{n-4}  \nonumber\\
 &\bleq - \frac{2 e \phi}{\hbar} \dot{S} a_{n-4} + \frac{2 \mathrm{i} e \phi}{\hbar} 
  \dot{a}_{n-6} - 2 e (\vec{A} \cdot \grad S) a_{n-2} 
  + 2 \mathrm{i} e \vec{A} \cdot \grad a_{n-4} \nonumber\\
 &\bleq - \frac{e^2 \phi^2}{\hbar^2} a_{n-6} + \frac{e^2 A^2}{\hbar} a_{n-4} 
  + m^2 a_{n-2} + \frac{\mathrm{i} e}{\hbar} \dot{\phi} a_{n-6} 
  + \mathrm{i} e (\grad \cdot \vec{A}) a_{n-4} \Big]
\end{align}
Sorting by powers of $n$ we now get the equations
\begin{align}
 0 &= \hbar (\grad S)^2 a_n \nonumber\\
  &\bleq + \left( m^2 - \dot{S}^2 - \mathrm{i} \hbar \Delta S 
  - 2 e (\vec{A} \cdot \grad S) \right) a_{n-2} - 2 \mathrm{i} \hbar (\grad S) 
  \cdot \grad a_{n-2} \nonumber\\
  &\bleq + \frac{1}{\hbar} \left(-\mathrm{i} \hbar \grad - e \vec{A}\right)^2 a_{n-4} 
  + \left( \mathrm{i} \ddot{S} - \frac{2 e \phi}{\hbar} \dot{S} \right) a_{n-4} 
  + 2 \mathrm{i} \dot{S} \dot{a}_{n-4} \nonumber\\
  &\bleq - \frac{1}{\hbar^2} \left( \mathrm{i} \hbar \partial_t 
  - e \phi \right)^2 a_{n-6}.
\end{align}

At order $n=0$ this yields simply $\grad S = 0$, thus 
$S(\vec{x},t)=S(t)$ depends only on time.

At order $n=2$ we then get
\begin{equation}
 (m^2 - \dot{S}^2) a_0 = 0 \quad \Rightarrow \quad S = \pm m t + \mathrm{const.}\,,
\end{equation}
where the constant term can be ignored and we choose the 
positive energy solution $S = -mt$.

Using these results at order $n=4$ finally yields the Schr\"odinger 
equation
\begin{equation}
 \left(\mathrm{i} \hbar \partial_t - e \phi\right) a_0 
  = \frac{1}{2 m} \left(-\mathrm{i} \hbar \grad - e \vec{A}\right)^2 a_0\,.
\end{equation}

At order $n=6$ we get
\begin{equation}
 \left(\mathrm{i} \hbar \partial_t - e \phi\right) a_2 = \frac{1}{2 m} 
  \left(-\mathrm{i} \hbar \grad - e \vec{A}\right)^2 a_2 
  - \frac{1}{2 m \hbar} \left( \mathrm{i} \hbar \partial_t - e \phi \right)^2 a_0.
\end{equation}
Neglecting the vector potential $\vec{A}$ this reduces to the
equation already found by Kiefer and Singh~\cite{Kiefer:1991}.
Without any electromagnetic potentials, we have
\begin{equation}
\mathrm{i} \hbar \dot{a}_2 = -\frac{\hbar^2}{2 m} \Delta a_2 
  - \frac{\hbar^3}{8 m^3} \Delta \Delta a_0.
\end{equation}

\subsection{Gravitating Klein-Gordon fields}
Next we consider a Klein-Gordon field coupled 
to Einstein's equations
\begin{equation}
 \label{eqn:Einstein-eq}
 G_{\mu \nu} = \frac{8 \pi \, G}{c^4} \, T_{\mu \nu},
\end{equation}
where
\begin{equation}
 G_{\mu \nu} = R_{\mu \nu} - \frac{1}{2} \, g_{\mu \nu} \, R
\end{equation}
is the Einstein tensor and $T_{\mu\nu}$ the energy-momentum tensor 
of the Klein-Gordon field, the expression of which will be given 
below (cf.~\eqref{eq:EM-TensorKG}). The set of equations 
\eqref{eqn:free-KG} and \eqref{eqn:Einstein-eq} is also known as 
the Einstein-Klein-Gordon system. We specialise to spherically-symmetric 
metrics which, upon choosing appropriate coordinates, we may write 
in the form~\cite{Straumann:GR}
\begin{equation}
\D s^2 = -\e^{2 A(r,t)} \, c^2 \, \D t^2 + \e^{2 B(r,t)} \, \D r^2 
  + r^2 \,(\D \theta^2 + \sin^2 \theta \, \D \varphi^2),
\end{equation}
with determinant
\begin{equation}
 g = -c^2 \e^{2(A+B)} r^4 \sin^2 \theta.
\end{equation}
We expand $\e^{A}$ and $\e^{B}$ as
\begin{align}
 \e^{A(r,t)} &= \sum_{n=0}^\infty \left(\frac{\sqrt{\hbar}}{c}\right)^n
                A_n(r,t) \quad ; \quad A_0 \equiv 1 \\
 \e^{B(r,t)} &= \sum_{n=0}^\infty \left(\frac{\sqrt{\hbar}}{c}\right)^n
                B_n(r,t) \quad ; \quad B_0 \equiv 1
\end{align}
and make use of the further expansions
{\allowdisplaybreaks%
\begin{equation}%
\label{eqn:exponentials-expansion}%
\begin{aligned}
\e^{-A} &= \sum_{n=0}^\infty \left(\frac{\sqrt{\hbar}}{c}\right)^n C_n, &
\e^{-B} - 1 &= \sum_{n=0}^\infty \left(\frac{\sqrt{\hbar}}{c}\right)^n
               D_n, \\
\e^{-A} \dot{B} &= \sum_{n=0}^\infty \left(\frac{\sqrt{\hbar}}{c}\right)^n
                    E_n, &
\e^{-B} A' &= \sum_{n=0}^\infty \left(\frac{\sqrt{\hbar}}{c}\right)^n
                    F_n, \\
\e^{-2 A} &= \sum_{n=0}^\infty \left(\frac{\sqrt{\hbar}}{c}\right)^n
                    G_n, &
\e^{-2 B} - 1 &= \sum_{n=0}^\infty \left(\frac{\sqrt{\hbar}}{c}\right)^n
                    H_n, \\
\e^{-2 A} (\dot{A} - \dot{B}) &= \sum_{n=0}^\infty
                          \left(\frac{\sqrt{\hbar}}{c}\right)^n J_n, &
\e^{-2 B} (A' - B') &= \sum_{n=0}^\infty
                   \left(\frac{\sqrt{\hbar}}{c}\right)^n K_n,
\end{aligned}%
\end{equation}}
with the coefficients
{\allowdisplaybreaks%
\begin{alignat}{3}
C_0 &= 1,\hspace{30pt} & C_1 &= -A_1,\hspace{30pt}      
    & C_2 &= A_1^2 - A_2 \nonumber\\
D_0 &= 0,\hspace{30pt} & D_1 &= -B_1,\hspace{30pt}
    & D_2 &= B_1^2 - B_2 \nonumber\\
E_0 &= 0,\hspace{30pt} & E_1 &= \dot{B}_1,\hspace{30pt}
    & E_2 &= -(A_1+B_1) \cdot{B}_1 + \dot{B}_2 \nonumber\\
F_0 &= 0,\hspace{30pt} & F_1 &= A_1',\hspace{30pt}
    & F_2 &= -(A_1+B_1) A_1' + A_2' \nonumber\\
G_0 &= 1,\hspace{30pt} & G_1 &= -2 A_1,\hspace{30pt}
    & G_2 &= 3 A_1^2 - 2 A_2 \nonumber\\
H_0 &= 0,\hspace{30pt} & H_1 &= -2 B_1,\hspace{30pt}
    & H_2 &= 3 B_1^2 - 2 B_2 \nonumber\\
J_0 &= 0,\hspace{30pt} & J_1 &= \dot{A}_1 - \dot{B}_1, && \nonumber\\
&& J_2 &= -2 A_1 (\dot{A}_1 - \dot{B}_1) - A_1 \dot{A}_1 
    + B_1 \dot{B}_1 + \dot{A}_2 - \dot{B}_2 \hspace{-200pt} && \nonumber\\
K_0 &= 0,\quad & K_1 &= A_1' - B_1', && \nonumber\\ 
&& K_2 &= -2 B_1 (A_1' - B_1') - A_1 A_1' + B_1 B_1' 
    + A_2' - B_2'\;. \hspace{-200pt} &&\nonumber
\end{alignat}}

The d'Alembert operator $\Box$ in a curved background is
\begin{align}
 \Box &= \frac{1}{\sqrt{-g}} \,\partial_\mu
  \left( \sqrt{-g} \, g^{\mu \nu} \,\partial_\nu \cdot \,\right) \\
 &= \e^{-2A} c^{-2} \left((\dot{A}-\dot{B}) \partial_t 
    - \partial_t^2\right) + \e^{-2B} (A'-B') \partial_r \nonumber\\
&\bleq + \left(\e^{-2B} - 1\right) \left(\frac{2}{r} \partial_r 
       + \partial_r^2\right) + \Delta
\end{align}
where the dot denotes derivatives with respect to $t$, the prime denotes
derivatives with respect to $r$ and $\Delta$ is the flat, three-dimensional
Laplace operator.

The Klein-Gordon-Equation then takes the following form:
\begin{align}
0 &= \e^{-2A} \,c^{-2} \, \left(\left(\dot{A} - \dot{B}\right) \, \dot{\psi}
    - \ddot{\psi}\right) + \e^{-2B} \, (A' - B') \, \psi' \nonumber\\
&\bleq + \left(\e^{-2B} - 1\right) \left(\frac{2}{r} \psi' + \psi'' \right)
       + \Delta \psi - \frac{m^2 \, c^2}{\hbar^2} \,\psi \,.
\end{align}

Using the same expansion \eqref{eqn:ansatz-fields} for $\psi$ as before,
and denoting by
$\Delta_r = 2/r \,\partial_r + \partial_r^2$ the radial component of the
spatial Laplacian, this becomes
\begin{align}
0 &= \exp\left(\frac{\mathrm{i} c^2}{\hbar} S\right) \frac{c^4}{\hbar^2}
  \sum_{n=0}^\infty \left(\frac{\sqrt{\hbar}}{c}\right)^n
  \Bigg[ \frac{1}{\hbar} \e^{-2A} \, \Big(\left(\dot{A} - \dot{B}\right)
  \left( \mathrm{i} \dot{S} a_{n-4} + \dot{a}_{n-6} \right) \nonumber\\
&\bleq + \dot{S}^2 a_{n-2} - 2 \mathrm{i} \dot{S} \dot{a}_{n-4}
  - \mathrm{i} \ddot{S} a_{n-4} - \ddot{a}_{n-6} \Big)
  + \e^{-2B} \, (A' - B') \left(\mathrm{i} S' a_{n-2} + a_{n-4}'\right) \nonumber\\
&\bleq + \left(\e^{-2B} - 1\right) \left(- S'^2 a_n + 2 \mathrm{i} S' a_{n-2}'
  + \mathrm{i} (\Delta_r S) a_{n-2} + \Delta_r a_{n-4} \right) \nonumber\\
&\bleq - (\grad S)^2 a_n + 2 \mathrm{i} (\grad S) \cdot \grad a_{n-2}
  + \mathrm{i} (\Delta S) a_{n-2} + \Delta a_{n-4}
  - \frac{m^2}{\hbar} a_{n-2} \Bigg]
\end{align}
and with the expansion for the exponentials
\eqref{eqn:exponentials-expansion} we obtain
\begin{align}
0 &= \exp\left(\frac{\mathrm{i} c^2}{\hbar} S\right) \frac{c^4}{\hbar^2}
  \sum_{n=0}^\infty \left(\frac{\sqrt{\hbar}}{c}\right)^n
  \Bigg[ \sum_{k=0}^n \Bigg\{ \frac{1}{\hbar} J_k \left( \mathrm{i} \dot{S} a_{n-k-4}
  + \dot{a}_{n-k-6} \right) \nonumber\\
&\bleq + \frac{1}{\hbar} G_k \left(\dot{S}^2 a_{n-k-2}
  - 2 \mathrm{i} \dot{S} \dot{a}_{n-k-4} - \mathrm{i} \ddot{S} a_{n-k-4}
  - \ddot{a}_{n-k-6} \right) \nonumber\\
&\bleq + K_k \left(\mathrm{i} S' a_{n-k-2} + a_{n-k-4}'\right) \nonumber\\
&\bleq + H_k \left(- S'^2 a_{n-k}  + 2 \mathrm{i} S' a_{n-k-2}'
  + \mathrm{i} (\Delta_r S) a_{n-k-2} + \Delta_r a_{n-k-4} \right) \Bigg\} \nonumber\\
&\bleq - (\grad S)^2 a_n + 2 \mathrm{i} (\grad S) \cdot \grad a_{n-2}
  + \mathrm{i} (\Delta S) a_{n-2} + \Delta a_{n-4} - \frac{m^2}{\hbar} a_{n-2} \Bigg].
\end{align}
As $H_0 = J_0 = K_0 = 0$ this can still be simplified and for each $n$ we get
\begin{align}
\label{eqn:KG-expansion}
0 &= \hbar (\grad S)^2 a_n + m^2 a_{n-2} - 2 \mathrm{i}\hbar  (\grad S)
  \cdot \grad a_{n-2}- \mathrm{i}\hbar  (\Delta S) a_{n-2}
  - \dot{S}^2 a_{n-2}  \nonumber\\
&\bleq + 2 \mathrm{i} \dot{S} \dot{a}_{n-4} + \mathrm{i} \ddot{S} a_{n-4}
  - \hbar \Delta a_{n-4} + \ddot{a}_{n-6} - \sum_{k=1}^n \Bigg[
  J_k \left( \mathrm{i} \dot{S} a_{n-k-4} + \dot{a}_{n-k-6} \right) \nonumber\\
&\bleq +  G_k \left(\dot{S}^2 a_{n-k-2} - 2 \mathrm{i} \dot{S} \dot{a}_{n-k-4}
  - \mathrm{i} \ddot{S} a_{n-k-4} - \ddot{a}_{n-k-6} \right) \nonumber\\
&\bleq + K_k \left(\mathrm{i}\hbar  S' a_{n-k-2} + \hbar a_{n-k-4}'\right) \nonumber\\
&\bleq + H_k \left(- \hbar S'^2 a_{n-k}  + 2 \mathrm{i} \hbar S' a_{n-k-2}' 
  + \mathrm{i} \hbar (\Delta_r S) a_{n-k-2} + \hbar \Delta_r a_{n-k-4} \right) \Bigg]\,.
\end{align}

At lowest order $n=0$ \eqref{eqn:KG-expansion} is again equivalent to
$(\grad S)^2 = 0$. Thus, $S$ is a function of
time only and \eqref{eqn:KG-expansion} is trivially fulfilled at order $n=1$.

At order $n=2$ we get $\dot{S}^2 = m^2$ and choose, as before, the
positive energy solution $S = - m t$. With these results equation
\eqref{eqn:KG-expansion} reduces to
\begin{align}
\label{eqn:KG-expansion-2}
0 &= -\mathrm{i} \hbar \dot{a}_{n-4} - \frac{\hbar^2}{2 m} \Delta a_{n-4} 
  + \frac{\hbar}{2 m} \ddot{a}_{n-6} \nonumber\\
&\bleq - \frac{\hbar}{2 m} \sum_{k=1}^n
  \Big[ G_k \left(m^2 a_{n-k-2} + 2 \mathrm{i} m \dot{a}_{n-k-4}
  - \ddot{a}_{n-k-6} \right) \nonumber\\
&\bleq + \hbar \left( K_k a_{n-k-4}' + H_k \Delta_r a_{n-k-4} \right)
  - J_k \left( \mathrm{i} m a_{n-k-4}
  - \dot{a}_{n-k-6} \right) \Big]\,.
\end{align}

At order $n=3$ we now obtain $G_1 = 0$ and therefore $A_1 = 0$, $G_2 = -2 A_2$.

Considering \eqref{eqn:KG-expansion} at order $n=4$, we finally see that the Klein-Gordon equation is equivalent to the \sne
\begin{equation}
 \mathrm{i} \hbar \dot{a}_0 = -\frac{\hbar^2}{2 m} \Delta a_0 + V a_0
\end{equation}
with potential $V = m \hbar A_2$.

\subsubsection{Einstein's equations}
Let us now consider Einstein's equations to derive the Poisson equation for
the potential $V$.

The non-vanishing components of the Einstein tensor are
\begin{subequations}%
\label{eqn:Einstein-tensor}%
\begin{align}
G_{tt}              &=  \e^{2A} \, c^2\, \left( \frac{1}{r^2} - \e^{-2B}\,
                        \left( \frac{1}{r^2} - \frac{2\,B'}{r} \right)\right)\\
G_{rr}              &=  -\frac{1}{r^2} \,\e^{2B} + \frac{1}{r^2} +
                        \frac{2\,A'}{r} \\
G_{tr}              &=  \frac{2\,\dot{B}}{r} \\
G_{\theta \theta}   &=  r^2 \,\e^{-2B} \left(A'^2 - A'\,B' + A''
                        + \frac{A'-B'}{r} \right) \nonumber\\
                    &\bleq + \frac{r^2}{c^2} \, \e^{-2A}
                        \,\left(-\dot{B}^2 + \dot{A}\dot{B} - \ddot{B}\right)\\
G_{\varphi \varphi} &=  \sin^2 \theta \, G_{\theta \theta}.
\end{align}%
\end{subequations}
From the Lagrangian for the Klein-Gordon field
\begin{equation}
 \mathcal{L} = -\frac{\hbar^2}{2m}
               \left((\partial^\lambda \psi)(\partial_\lambda
\psi^*) + \frac{m^2 c^2}{\hbar^2}\,\abs{\psi}^2\right) \sqrt{-g}
\end{equation}
the stress-energy-tensor can be derived as
\begin{align}
\label{eq:EM-TensorKG}
T_{\mu \nu} &= -\frac{2}{\sqrt{-g}}
               \frac{\delta \mathcal{L}}{\delta g^{\mu \nu}} \nonumber\\
&= \frac{\hbar^2}{2m} \,\Bigg[
  (\partial_\mu \psi)(\partial_\nu \psi^*)
  + (\partial_\mu \psi^*)(\partial_\nu \psi)
  - \,g_{\mu \nu} \left((\partial^\lambda \psi)(\partial_\lambda \psi^*)
  + \frac{m^2 c^2}{\hbar^2} \,\abs{\psi}^2 \right) \Bigg].
\end{align}
Its non-vanishing components are
\begin{subequations}%
\begin{align}
T_{tt}              &=  \frac{m c^4}{2}\,\e^{2A}\,\abs{\psi}^2
                        + \frac{\hbar^2 c^2}{2m}
                        \,\e^{2(A-B)} \abs{\psi'}^2
                        + \frac{\hbar^2}{2m}\,\abs{\dot{\psi}}^2\\
T_{rr}              &=  - \frac{m c^2}{2}\,\e^{2B}\,\abs{\psi}^2
                        + \frac{\hbar^2}{2m}\,\abs{\psi'}^2
                        + \frac{\hbar^2}{2m\,c^2}
                        \,\e^{2(B-A)} \abs{\dot{\psi}}^2\\
T_{tr}              &=  \frac{\hbar^2}{2m}\,(\dot{\psi} \psi'^* +
                        \dot{\psi}^* \psi')\\
T_{\theta \theta}   &=  - \frac{m c^2}{2} r^2\,\abs{\psi}^2
                        - \frac{ \hbar^2\,r^2}{2m}\,\e^{-2B}\abs{\psi'}^2
                        + \frac{ \hbar^2\,r^2}{2m\,c^2}
                        \,\e^{-2A}\,\abs{\dot{\psi}}^2\\
T_{\varphi \varphi} &=  \sin^2 \theta \, T_{\theta \theta}.
\end{align}%
\end{subequations}

We now expand both the Einstein tensor $G_{\mu \nu}$ and $T_{\mu \nu} / c^4$
using \emph{Mathematica} and
consider Einstein's equations for each component order by order. We make use of
the fact that $A_1 = 0$ from our analysis of the Klein-Gordon
equation, and we use the lower order results to simplify the equations at
higher order. The components that are not mentioned are trivially fulfilled at
the given order.

\paragraph{\boldmath$n=0$:}
The $tt$-component yields $(r B_1)' = 0$.

\paragraph{\boldmath$n=1$:}
The $tt$-component yields
\begin{equation}
\label{eqn:KG-n1-tt}
 \frac{2 \hbar}{r^2} \left(\frac{3}{2} B_1^2 + (r B_2)' \right)
  = 8 \pi G m \abs{a_0}^2.
\end{equation}

\paragraph{\boldmath$n=2$:}
The $tt$-component yields
\begin{equation}
\label{eqn:KG-n2-tt}
 \frac{2 \hbar}{r^2} \left(-4 B_1^3 - 3 r B_1 B_2' + (r B_3)' \right)
  = 8 \pi G m \left(a_1^* a_0 + a_0^* a_1 \right).
\end{equation}
The $rr$- and $\theta \theta$-components both yield $B_1=0$ and the
$tr$-component $B_1'=0$ is then trivial as well as the order $n=0$ equation.
Equations \eqref{eqn:KG-n1-tt} and \eqref{eqn:KG-n2-tt} then simplify to
\begin{align}
(r B_2)' &= \frac{4 \pi G m r^2}{\hbar} \abs{a_0}^2 \\
\label{eqn:KG-n2-tt-simplified}
(r B_3)' &= \frac{4 \pi G m r^2}{\hbar} \left(a_1 a_0^* + a_0 a_1^* \right)
\end{align}

\paragraph{\boldmath$n=3$:}
The $tt$-component yields
\begin{multline}
\label{eqn:KG-n3-tt}
4 (A_2-B_2) (r B_2)' + B_2^2 - 2 r B_2 B_2' + (r B_4)'
  = \frac{4 \pi G r^2}{\hbar^2} \Big[ \frac{\hbar^2}{2 m}\abs{a_0'}^2 \\
 + \frac{\mathrm{i} \hbar}{2} \left( a_0^* \dot{a}_0 - \dot{a}_0^* a_0 \right)
  + m \hbar \left( A_2 \abs{a_0}^2 + \abs{a_1}^2 + a_2^* a_0
  + a_0^* a_2 \right) \Big].
\end{multline}
The $rr$-component yields
\begin{equation}
 B_2 = r A_2'
\end{equation}
and the $\theta \theta$-component is just the derivative of the $rr$-component.
The $tr$-component yields
\begin{equation}
\frac{2}{r} \dot{B}_2 = 4 \pi \mathrm{i} G \left(a_0^* a_0' - a_0 {a_0^*}' \right).
\end{equation}

If we define the potential $V = m \hbar A_2$ as before and analogously
$U = m \hbar B_2$, and also introduce the ($r$-component of the) 
probability current
\begin{equation}
 j_\text{KG} = \frac{\mathrm{i} \hbar}{2 m} \left( {a_0^*}' a_0 - a_0^* a_0' \right),
\end{equation}
we are left with the following set of equations:
\begin{subequations}%
\label{eqn:set-pre-poisson}
\begin{align}
\label{eqn:final-set-U}
U &= r V' \\
\label{eqn:final-set-rU}
(r U)' &= 4 \pi G m^2 r^2 \abs{a_0}^2 \\
\label{eqn:final-set-dotU}
\dot{U} &= -4 \pi G m^2 r j,
\end{align}%
\end{subequations}
which are equations for $U$, $V$ and $a_0$ only, together with equations
\eqref{eqn:KG-n2-tt-simplified} and \eqref{eqn:KG-n3-tt} which constrain
$B_3$, $B_4$ and $a_2$ in terms of $U$, $V$ and $a_0$.

\subsubsection{Poisson equation} \label{sec:Poisson}
Let us further analyse the set of equations \eqref{eqn:set-pre-poisson} to show
that they are equivalent to the Poisson equation for the potential $V$.

$U$ is determined from \eqref{eqn:final-set-rU} to be
\begin{equation}
\label{eqn:U-result}
U(r,t) = \frac{4 \pi G \,m^2}{r} \,\int_0^r \D \widetilde{r} \, \widetilde{r}^2
\,\abs{a_0(\widetilde{r},t)}^2.
\end{equation}
Inserting \eqref{eqn:final-set-U} into \eqref{eqn:final-set-rU} yields
\begin{equation}
4 \pi \, G \, m^2 \, \abs{a_0}^2 = V'' + \frac{2}{r} V'
= \Delta V
\end{equation}
and therefore the Laplace equation for $V$ that we were looking for.

We still have to check the consistency of equation \eqref{eqn:final-set-dotU}.
Note that by the continuity equation which directly follows from the Schr\"odinger
equation we get
\begin{equation}
\mathrm{div} \vec{j} + \partial_t \abs{a_0}^2 = 0,
\end{equation}
where in the spherically symmetric case the divergence is given by
\begin{equation}
\mathrm{div} \vec{j} = j' + \frac{2}{r} \, j.
\end{equation}
Differentiating \eqref{eqn:U-result} by $t$ then yields
\begin{align}
\dot{U} &= 4\pi \,G\,m^2\,r \int_0^r \D \widetilde{r}\,
          \frac{\widetilde{r}^2}{r^2} \, \partial_t
          \abs{a_0(\widetilde{r},t)}^2 \nonumber\\
        &= -4\pi \,G\,m^2\,r \,\int_0^1 \D x\, x^2 \mathrm{div}
          \vec{j}(xr,t) \nonumber\\
        &= -G\,m^2\,r \, \int_{\text{unit sphere}} \hspace{-30pt} 
          \D \mathrm{V}\; \mathrm{div}\vec{j}(xr,t)
          \nonumber\\
        &= - \,G\,m^2\,r\,\int_{\partial(\text{unit sphere})}
          \hspace{-40pt} \D \mathrm{S}\; j(r,t) \nonumber\\
        &= -4\pi \,G\,m^2\,r\, j.
\end{align}
Hence, equation \eqref{eqn:final-set-dotU} already follows from the other
equations and the system is consistent.

\section{Dirac fields} \label{par:dirac}
Let us now turn to the Dirac equation and repeat our analysis for this case.
The free Dirac equation is
\begin{equation}
\label{eqn:free_dirac_1}
 \left( \mathrm{i} \gamma^\mu \partial_\mu - \frac{m c}{\hbar} \right) \psi = 0,
 \hspace{1cm} \gamma^0 = \tmatrix{\1}{0}{0}{-\1},\;
 \gamma^k = \tmatrix{0}{\sigma^k}{-\sigma^k}{0}
\end{equation}
with the Pauli matrices
\begin{equation}
 \sigma^1 = \tmatrix{0}{1}{1}{0}, \hspace{1cm}
  \sigma^2 = \tmatrix{0}{-\mathrm{i}}{\mathrm{i}}{0}, \hspace{1cm}
  \sigma^3 = \tmatrix{1}{0}{0}{-1}.
\end{equation}
$\psi$ here is a four component spinor field.

Defining the matrices $\beta = \gamma^0$ and $\alpha^k = \gamma^0 \gamma^k$
and multiplying equation \eqref{eqn:free_dirac_1} by $\beta c$ yields
\begin{equation}
 \left( \mathrm{i} \partial_t + \mathrm{i} c \alpha^k \partial_k
 - \frac{m c^2}{\hbar} \beta \right) \psi = 0.
\end{equation}

Now we can, again, introduce an electromagnetic field with electric
potential $\phi$ and vector potential $\vec{A}$ by replacing
\begin{alignat}{2}
 \partial_t &\rightarrow \partial_t &&+ \frac{\mathrm{i} e}{\hbar} \phi(\vec{x},t) \\
 \partial_k &\rightarrow \partial_k &&- \frac{\mathrm{i} e}{\hbar} A_k(\vec{x},t).
\end{alignat}
The Dirac equation then takes the form
\begin{equation}
\label{eqn:dirac}
 \left( \mathrm{i} \partial_t +\mathrm{i} c \alpha^k \partial_k - \frac{e}{\hbar} \phi
  + \frac{e c}{\hbar} \alpha^k A_k
  - \frac{m c^2}{\hbar} \beta \right) \psi = 0.
\end{equation}

As for the Klein-Gordon equation, we now make use of our ansatz
\eqref{eqn:ansatz-fields} with the derivatives \eqref{eqn:ansatz-derivatives},
where $S$ now is a scalar function but the $a_n$ are four component spinors.
Inserting this ansatz into the Dirac equation \eqref{eqn:dirac} yields
\begin{align}
\label{eqn:ansatz-dirac}
 0 &= \exp\left(\frac{\mathrm{i} c^2}{\hbar} S\right) \frac{c^2}{\hbar}
  \sum_{n=0}^\infty \left(\frac{\sqrt{\hbar}}{c}\right)^n
  \Big[ -\dot{S} a_n + \mathrm{i} \dot{a}_{n-2}
  - \frac{e \phi}{\hbar} a_{n-2} \nonumber\\
&\bleq - c \vec{\alpha} \cdot (\grad S) a_n
  + \mathrm{i} c \vec{\alpha} \cdot \grad a_{n-2}
  + \frac{e c}{\hbar} \vec{\alpha} \cdot \vec{A} a_{n-2}
  - \frac{m c^2}{\hbar} \beta a_{n-2} \Big].
\end{align}

\subsection{The semi-classical limit}
As for the \kg we rewrite equation \eqref{eqn:ansatz-dirac}
eliminating the $\hbar$ terms:
\begin{align}
 0 &= \exp\left(\frac{\mathrm{i} c^2}{\hbar} S\right) \frac{c^2}{\hbar}
  \sum_{n=0}^\infty \left(\frac{\sqrt{\hbar}}{c}\right)^n
  \Big[ -\dot{S} a_n + \mathrm{i} \dot{a}_{n-2}
  - \frac{e \phi}{c^2} a_{n} \nonumber\\
&\bleq - c \vec{\alpha} \cdot (\grad S) a_n
  + \mathrm{i} c \vec{\alpha} \cdot \grad a_{n-2}
  + \frac{e}{c} \vec{\alpha} \cdot \vec{A} a_{n} - m \beta a_{n} \Big].
\end{align}
Sorting by powers of $n$ this yields
\begin{equation}
 \left( m \beta + \dot{S} + \frac{e \phi}{c^2} + c \vec{\alpha} \cdot \grad S
  - \frac{e}{c} \vec{\alpha} \cdot \vec{A} \right) a_n - \mathrm{i} \dot{a}_{n-2}
  - \mathrm{i} c \vec{\alpha} \cdot \grad a_{n-2} = 0\,.
\end{equation}

Making use of the notations $\pi_0 = -c \dot{S} - e \phi / c$ and
$\pi_k = -c^2 \partial_k S + e A_k$ we obtain at order $n=0$
\begin{align}
\label{eqn:dirac-class-lim-first-order}
0 &= \left( m c \beta - \pi_0 - \vec{\alpha}
  \cdot \vec{\pi} \right) a_0 \nonumber\\
\Leftrightarrow \quad 0 &= \left( m c - \pi_\mu \gamma^\mu \right)
  a_0 \nonumber\\
  &= \left(\begin{array}{cc} (m c - \pi_0) \1 & -\vec{\sigma}
  \cdot \vec{\pi} \\ \vec{\sigma} \cdot \vec{\pi}
  & (m c + \pi_0) \1 \end{array}\right) a_0 ,
\end{align}
which has non-trivial solutions if and only if the determinant
\begin{equation}
\label{eqn:vanishing-determinant-condition}
 \left|\begin{array}{cc} (m c - \pi_0) \1 & -\vec{\sigma} \cdot \vec{\pi} \\
 \vec{\sigma} \cdot \vec{\pi} & (m c + \pi_0) \1 \end{array}\right|
  = m^2 c^2 - \pi_0^2 + \left| (\vec{\sigma} \cdot \vec{\pi})^2 \right| = 0
\end{equation}
vanishes. The Pauli matrices obey the algebra
\begin{equation}
\label{eqn:Pauli-matrices-algebra}
 \sigma^i \sigma^j = \delta^{ij} + \mathrm{i} \epsilon^{ijk} \sigma^k \quad
  \Rightarrow \quad
 (\sigma \cdot \vec{u}) (\sigma \cdot \vec{v})
  = \vec{u} \cdot \vec{v} + \mathrm{i} \vec{\sigma} \cdot (\vec{u} \times \vec{v}),
\end{equation}
therefore $(\vec{\sigma} \cdot \vec{\pi})^2 = \vec{\pi} \cdot \vec{\pi}$ and 
\eqref{eqn:vanishing-determinant-condition} yields again, as for the \kg,
the Hamilton-Jacobi equation for a relativistic particle
\begin{equation}
 0 = m^2 c^2 + \pi_\mu \pi^\mu.
\end{equation}

At order $n=2$ we get
\begin{align}
\label{eqn:class-limit-dirac-second-order}
 \left( m c - \pi_\mu \gamma^\mu \right) a_2 &=
  \mathrm{i} c \gamma^0 \left(\partial_t
  + c \vec{\alpha} \cdot \grad \right) a_0 \nonumber\\
 &= -\mathrm{i} c^2 \gamma^\mu \partial_\mu a_0.
\end{align}

If we name the operators by
\begin{equation}
 L :=  m c - \pi_\mu \gamma^\mu \;, \quad
 D := -\mathrm{i} c^2 \gamma^\mu \partial_\mu \;,
\end{equation}
at order $n=0$ we have the condition that $a_0 \in \mathrm{Ker}(L)$. At second
order we now have $D a_0 \in \mathrm{Im}(L)$ which is equivalent to
$D a_0 \in \left(\mathrm{Ker} (L^\dagger)\right)^\perp$.
Now note that $L^\dagger = \gamma^0 L \gamma^0$ and therefore
$x \in \mathrm{Ker} (L^\dagger)
\Leftrightarrow \gamma^0 x \in \mathrm{Ker} (L)$.

The condition that $D a_0$ is in the image of $L$ is therefore equivalent to the
condition that for any two solutions $a_0, \widetilde{a}_0 \in \mathrm{Ker} (L)$
to the first order equation we have
\begin{equation}
\label{eqn:condition-for-a0-at-second-order}
 \overline{a}_0 \gamma^\mu \partial_\mu \widetilde{a}_0 = 0,
\end{equation}
where $\overline{a}_0 = (a_0)^\dagger \gamma_0$ is the adjoint spinor.
Equations \eqref{eqn:dirac-class-lim-first-order}
and \eqref{eqn:condition-for-a0-at-second-order} together determine
the solutions $a_0$ at first order.

\subsubsection{Derivation of the BMT equation}
We can use these results to obtain the Bargmann-Michel-Telegdi
equation~\cite{Bargmann:1959} as a necessary condition. First multiply
equation \eqref{eqn:class-limit-dirac-second-order} by
$(m c + \pi_\mu \gamma^\mu)$ from the left.
Then the left-hand side vanishes and we get
\begin{equation}
\label{eqn:dirac-class-lim-second-order}
 (m c + \pi_\mu \gamma^\mu) \gamma^\nu \partial_\nu a_0 = 0.
\end{equation}
Now let us calculate
\begin{align}
\label{eqn:pimupartiala}
 -2 \pi^\mu \partial_\mu a_0 &= \pi_\mu (-2 \eta^{\mu \nu})
  \partial_\nu a_0 \nonumber\\
&= \pi_\mu \gamma^\mu \gamma^\nu \partial_\nu a_0
  + \pi_\mu \gamma^\nu \gamma^\mu \partial_\nu a_0 \nonumber\\
&= -m c \gamma^\nu \partial_\nu a_0
  + \partial_\nu \left( \gamma^\nu \pi_\mu \gamma^\mu a_0 \right)
  - (\partial_\nu \pi_\mu) \gamma^\nu \gamma^\mu a_0 \nonumber\\
&= - (\gamma^\nu \partial_\nu  \gamma^\mu \pi_\mu) a_0 \nonumber\\
&= \left( \partial_\mu \pi^\mu - \vec{\alpha} \cdot
  \left(\frac{1}{c} \dot{\vec{\pi}} - \grad \pi_0 \right)
  + \mathrm{i} \gamma^5 \vec{\alpha} \cdot \left(\grad \times \vec{\pi}\right) \right)
  a_0 \nonumber\\
&= (\partial_\mu \pi^\mu) a_0
  - e \left( \frac{1}{c} \vec{\alpha} \cdot \left(\vec{E}-\dot{\vec{A}}\right)
  + \mathrm{i} \gamma^5 \vec{\alpha} \cdot \vec{B}\right) a_0 \nonumber\\
&= (\partial_\mu \pi^\mu) a_0 + \frac{e}{2}
    F_{\mu \nu} \gamma^\mu \gamma^\nu a_0,
\end{align}
where $\gamma^5 = \mathrm{i} \gamma^0 \gamma^1 \gamma^2 \gamma^3$ and we made use of
\eqref{eqn:dirac-class-lim-second-order} in the third and
\eqref{eqn:dirac-class-lim-first-order} in the fourth line. Performing the
same calculation for the adjoint leads to
\begin{equation}
\label{eqn:pimupartiala-adj}
 -2 \pi^\mu \partial_\mu \overline{a}_0
  = (\partial_\mu \pi^\mu) \overline{a}_0
  - \frac{e}{2} F_{\mu \nu} \overline{a}_0 \gamma^\mu \gamma^\nu.
\end{equation}

We follow \cite{Rafanelli:1964} and set
\begin{equation}
 S^\mu = \overline{a}_0 \gamma^5 \gamma^\mu a_0,
\end{equation}
the spin density, then we
get from equations \eqref{eqn:pimupartiala} and \eqref{eqn:pimupartiala-adj}
\begin{align}
 \pi^\mu \partial_\mu S^\nu &= \pi^\mu (\partial_\mu \overline{a}_0)
  \gamma^5 \gamma^\nu a_0 + \overline{a}_0 \gamma^5 \gamma^\nu
  \pi^\mu \partial_\mu a_0 \nonumber\\
&= -(\partial_\mu \pi^\mu) S^\nu + \frac{e}{4} F_{\rho \sigma}
  \overline{a}_0 \left( \gamma^\rho \gamma^\sigma \gamma^5 \gamma^\nu
  - \gamma^5 \gamma^\nu \gamma^\rho \gamma^\sigma \right) a_0 \nonumber\\
&= -(\partial_\mu \pi^\mu) S^\nu + \frac{e}{4} F_{\rho \sigma}
  \overline{a}_0 \gamma^5 \left( 2 \eta^{\rho \nu} \gamma^\sigma
  - 2 \eta^{\sigma \nu} \gamma^\rho \right) a_0 \nonumber\\
&= -(\partial_\mu \pi^\mu) S^\nu + e F^\nu{}_\rho S^\rho.
\end{align}

For the normalised spin density
\begin{equation}
 \hat{S}^\mu = \frac{1}{\sqrt{-S_\nu S^\nu}} S^\mu
\end{equation}
the first term vanishes and we obtain the BMT equation (for g-factor $g=2$)
\begin{equation}
 \frac{\D}{\D \tau}\hat{S}^\nu
  =  \frac{1}{m} \pi^\mu \partial_\mu \hat{S}^\nu
  = \frac{e}{m} F^\nu{}_\rho \hat{S}^\rho,
\end{equation}
where the first equality holds because $\pi^\mu = m v^\mu$ with $v^\mu$
the four-velocity of the relativistic particle in an electromagnetic field.

\subsection{The non-relativistic limit} \label{sec:dirac-nr}
Again, we rewrite equation \eqref{eqn:ansatz-dirac} eliminating the $c$ terms:
\begin{align}
 0 &= \exp\left(\frac{\mathrm{i} c^2}{\hbar} S\right) \frac{c^3}{\hbar^{3/2}} 
  \sum_{n=0}^\infty \left(\frac{\sqrt{\hbar}}{c}\right)^n
  \Big[ -\dot{S} a_{n-1} + \mathrm{i} \dot{a}_{n-3}
  - \frac{e \phi}{\hbar} a_{n-3} \nonumber\\
&\bleq - \sqrt{\hbar} \vec{\alpha} \cdot (\grad S) a_{n}
  + \mathrm{i} \sqrt{\hbar} \vec{\alpha} \cdot \grad a_{n-2}
  + \frac{e}{\sqrt{\hbar}} \vec{\alpha} \cdot \vec{A} a_{n-2}
  - m \beta a_{n-1} \Big].
\end{align}
Sorting by powers of $n$ we get
\begin{multline}
\label{eqn:dirac_sorted}
 \sqrt{\hbar} \vec{\alpha} \cdot (\grad S) a_n + \left(\dot{S}
  + m \beta \right) a_{n-1} + \frac{1}{\sqrt{\hbar}} \vec{\alpha}
  \cdot \left(-\mathrm{i} \hbar \grad - e \vec{A}\right) a_{n-2} \\
  - \frac{1}{\hbar} \left(\mathrm{i} \hbar \partial_t - e \phi\right) a_{n-3} = 0.
\end{multline}

At order $n=0$ this simply becomes $\grad S = 0$, and therefore $S=S(t)$ is a
function of time only.

Now, splitting the four component spinors
$a_n = (a_{n,1}, a_{n,2}, a_{n,3}, a_{n,4})$
into two two component spinors
$a_n^>  = (a_{n,1}, a_{n,2})$ and $a_n^<  = (a_{n,3}, a_{n,4})$,
at order $n=1$ we get the two equations
\begin{subequations}\begin{align}
 (m + \dot{S}) a_0^> &= 0 \\
 (m - \dot{S}) a_0^< &= 0,
\end{align}\end{subequations}
which can be consistently fulfilled only if either $S=-mt$ and the
negative energy component $a_0^<$ vanishes or $S=+mt$ and the positive
energy component $a_0^>$ vanishes. From here on, we will choose the
first and set $a_0^< \equiv 0$. With this choice, the Dirac equation
\eqref{eqn:dirac_sorted} simplifies to the following two equations:
\begin{subequations}\begin{align}
\label{eqn:dirac-simplified-1}
 0 &= \sqrt{\hbar} \vec{\sigma} \cdot \left(-\mathrm{i} \hbar \grad 
  - e \vec{A}\right) a_{n-2}^< - \left(\mathrm{i} \hbar \partial_t 
  - e \phi\right) a_{n-3}^> \\
\label{eqn:dirac-simplified-2}
 a_{n-1}^< &= \frac{1}{2 m \sqrt{\hbar}} \vec{\sigma}
  \cdot \left(-\mathrm{i} \hbar \grad - e \vec{A}\right) a_{n-2}^>
  - \frac{1}{2 m} \left(\mathrm{i} \hbar \partial_t - e \phi\right) a_{n-3}^< \,.
\end{align}\end{subequations}

At order $n=2$ \eqref{eqn:dirac-simplified-1} is trivially fulfilled.
Equation \eqref{eqn:dirac-simplified-2} determines $a_1^<$ to be
\begin{equation}
\label{eqn:a1minus}
 a_{1}^< = \frac{1}{2 m \sqrt{\hbar}} \vec{\sigma} \cdot
  \left(-\mathrm{i} \hbar \grad - e \vec{A}\right) a_{0}^>.
\end{equation}

Using this at order $n=3$ in equation \eqref{eqn:dirac-simplified-1}, we get
\begin{equation}
 \left( \frac{1}{2 m} \left[\vec{\sigma} \cdot \left(-\mathrm{i} \hbar \grad
  - e \vec{A}\right)\right]^2 - \left(\mathrm{i} \hbar \partial_t
  - e \phi\right) \right) a_0^> = 0.
\end{equation}
Making use of
\begin{multline}
 \left(-\mathrm{i} \hbar \grad - e \vec{A}\right) \times \left(-\mathrm{i} \hbar \grad
  - e \vec{A}\right) a_0^> = -\hbar^2
  \underbrace{\grad \times (\grad a_0^>)}_{=0} \\
+ \mathrm{i} e \hbar \underbrace{\grad \times (a_0^> \vec{A})}_{=a_0^>
  \grad \times \vec{A} - \vec{A} \times \grad a_0^>}
+ \mathrm{i} e \hbar \vec{A} \times \grad a_0^> + e^2
  \underbrace{\vec{A} \times \vec{A}}_{=0} = \mathrm{i} e \hbar \vec{B} a_0^>
\end{multline}
and the algebra \eqref{eqn:Pauli-matrices-algebra} of the Pauli matrices
we end up with the Pauli equation
\begin{equation}
 \left(\mathrm{i} \hbar \partial_t - e \phi \right) a_0^> = \frac{1}{2 m}
  \left(-\mathrm{i} \hbar \grad - e \vec{A}\right)^2 a_0^>
  - \frac{e \hbar}{2 m} \vec{\sigma} \cdot \vec{B} a_0^>.
\end{equation}
Equation \eqref{eqn:dirac-simplified-2} at order $n=3$ determines $a_2^<$ to be
\begin{equation}
 a_{2}^< = \frac{1}{2 m \sqrt{\hbar}} \vec{\sigma}
  \cdot \left(-\mathrm{i} \hbar \grad - e \vec{A}\right) a_{1}^>.
\end{equation}
This is exactly the same relation as \eqref{eqn:a1minus} with the indices
shifted by one. Thus, from equation \eqref{eqn:dirac-simplified-1} at order
$n=4$ we will get the Pauli equation again for $a_1^>$.
But equation \eqref{eqn:dirac-simplified-2} at order $n=4$ now has an
additional term depending on $a_1^<$
\begin{equation}
 a_{3}^< = \frac{1}{2 m \sqrt{\hbar}} \vec{\sigma} \cdot
    \left(-\mathrm{i} \hbar \grad - e \vec{A}\right) a_{2}^>
    + \frac{1}{2 m} \left(-\mathrm{i} \hbar \partial_t + e \phi\right) a_{1}^<.
\end{equation}
This will make the equation for $a_2^>$ more complicated, and different from the
Pauli equation, at order $n=5$.

\subsection{Gravitating Dirac fields}
Now let us consider the Einstein-Dirac system, consisting of the
Dirac equation \eqref{eqn:free_dirac_1} and Einstein's
equations \eqref{eqn:Einstein-eq}.

As for the Einstein-Klein-Gordon system, we make an ansatz using
a spherically symmetric metric tensor and therefore also need the
stress-energy-tensor to be spherically symmetric.
Note that a single Dirac particle cannot have spherical symmetry,
which is why we have to average over all spin directions. But we
will only take this into account at the very end of our considerations.

We use the general relativistic formulation of the Dirac equation according
to \citet{Finster:1998} and \citet{Finster:1999}
\begin{equation}
\label{eqn:the-dirac-equation}
 \left( \slashed{D} - \frac{m c}{\hbar} \right) \psi = 0.
\end{equation}
The Dirac operator is defined as
\begin{subequations}%
\begin{align}
\slashed{D} &= \mathrm{i} \Gamma^\mu(x) \partial_\mu + Y(x) \\
Y(x) &= \Gamma^\mu(x) Z_\mu(x) \\
Z_\mu &= \frac{\mathrm{i}}{2} \rho \partial_\mu \rho
  - \frac{\mathrm{i}}{16} \tr{\Gamma^\nu \nabla_\mu \Gamma^\rho}
  \Gamma_\nu \Gamma_\rho
  + \frac{\mathrm{i}}{8} \tr{ \rho \Gamma_\mu \nabla_\nu \Gamma^\nu } \rho \\
\rho &= \frac{\mathrm{i}}{4!} \epsilon_{\mu \nu \rho \sigma}
  \Gamma^\mu \Gamma^\nu \Gamma^\rho \Gamma^\sigma,
\end{align}%
\end{subequations}
where $\nabla$ is the covariant derivative, $\epsilon_{\mu \nu \rho \sigma}$
is the Levi-Civita symbol defined by
\begin{equation}
 \epsilon_{\mu \nu \rho \sigma} = \sqrt{-g} \begin{cases} +1 &\mbox{if }
  (\mu \nu \rho \sigma) \mbox{ is an even permutation of } (t r \theta \varphi)
  \\ -1 &\mbox{if } (\mu \nu \rho \sigma) \mbox{ is an odd permutation of }
  (t r \theta \varphi) \\ 0 &\mbox{if two or more indices are equal}
  \end{cases}
\end{equation}
and $\Gamma^\mu$ is a representation of the space-time dependent Dirac
matrices, satisfying the Clifford algebra
\begin{equation}
 \{ \Gamma^\mu, \Gamma^\nu \} = -2 g_{\mu \nu}.
\end{equation}
The additional part $Y(x)$ will turn out not to contribute to the \sne and
is only relevant at higher order in $1/c$.

It is useful to represent these matrices in the basis where they become the
linear combination
\begin{align}
\label{eqn:choice-dirac-matrices}
\Gamma^t &= \e^{-A} c^{-1} \gamma^0 \\
\Gamma^r &= \e^{-B} \left( \gamma^1 \cos \theta
  + \gamma^2 \sin \theta \cos \varphi 
  + \gamma^3 \sin \theta \sin \varphi \right) \\
\Gamma^\theta &= \frac{1}{r} \left( -\gamma^1 \sin \theta 
  + \gamma^2 \cos \theta \cos \varphi 
  + \gamma^3 \cos \theta \sin \varphi \right) \\
\Gamma^\varphi &= \frac{1}{r \sin \theta} \left( - \gamma^2 \sin \varphi 
  + \gamma^3 \cos \varphi \right),
\end{align}
of the Dirac matrices $\gamma^\mu$ as defined before. These matrices satisfy
the anti-commutator algebra, and simplify the equations because in this 
representation $\rho = \gamma^5$ and therefore the first term of $Z_\mu$ 
vanishes because $\rho$ is constant, and the third term vanishes because 
derivatives of the $\Gamma^\mu$ as well as the $\Gamma^\mu$ themselves are 
linear in the $\gamma^\mu$ and $\tr{\gamma^5 \gamma^\mu \gamma^\nu} = 0$. 
Therefore
\begin{align}
 Y &= -\frac{\mathrm{i}}{16} \tr{\Gamma^\nu \nabla_\mu \Gamma^\rho} \Gamma^\mu 
  \Gamma_\nu \Gamma_\rho \\
&= -\frac{\mathrm{i}}{16} \tr{\Gamma^\nu \nabla_\mu \Gamma^\rho} \left(\delta_\rho^\mu
  \Gamma_\nu - \delta_\nu^\mu \Gamma_\rho - \Gamma^\mu g_{\nu \rho}
  + \mathrm{i} \epsilon^\mu{}_{\nu \rho \sigma} \Gamma^\sigma \gamma^5 \right) \\
&= -\frac{\mathrm{i}}{8} \tr{\Gamma^\nu \nabla_\rho \Gamma^\rho} \Gamma_\nu
  + \frac{1}{16} \epsilon^{\mu \nu \rho \sigma}
  \tr{\Gamma_\nu \nabla_\mu \Gamma_\rho} \Gamma_\sigma \gamma^5 \\
&= -\frac{\mathrm{i}}{8} \tr{\Gamma^\nu \nabla_\rho \Gamma^\rho} \Gamma_\nu.
\end{align}
All this is in agreement with \cite{Finster:1999} and can be straightforwardly
verified. The third line follows, because the $\delta^\mu_\nu$ terms in the
second line are equal and the $g_{\nu \rho}$ term vanishes. The second term
in the third line vanishes, because the trace vanishes if all three indices
are different.

Using that $\nabla_\rho \Gamma^\rho = \alpha_\rho \Gamma^\rho$ is some linear
combination of the gamma matrices we get
\begin{align}
 \tr{\Gamma^\nu \nabla_\rho \Gamma^\rho}
  &= \alpha_\rho \tr{\Gamma^\nu \Gamma^\rho} \\
&= \frac{1}{2} \alpha_\rho \tr{\{\Gamma^\nu, \Gamma^\rho\}
  + [\Gamma^\nu, \Gamma^\rho]} \\
&= \frac{1}{2} \alpha_\rho \left( \tr{\{\Gamma^\nu, \Gamma^\rho\}}
  + \tr{\Gamma^\nu \Gamma^\rho} - \tr{\Gamma^\rho \Gamma^\nu} \right) \\
&= \frac{1}{2} \alpha_\rho \left( \tr{-2 g^{\nu \rho} \1_{4 \times 4}}
  + \tr{\Gamma^\nu \Gamma^\rho} - \tr{\Gamma^\nu \Gamma^\rho} \right) \\
&= - \alpha_\rho g^{\nu \rho} \tr{\1_{4 \times 4}} \\
&= -4 \alpha^\nu
\end{align}
and therefore
\begin{equation}
 \slashed{D} = \mathrm{i} \Gamma^\mu \partial_\mu + \frac{\mathrm{i}}{2} \nabla_\mu \Gamma^\mu.
\end{equation}

For our ansatz for the Dirac matrices we get
\begin{align}
 \nabla_\mu \Gamma^\mu &= \frac{1}{\sqrt{-g}} \partial_\mu
  \left( \sqrt{-g} \; \Gamma^\mu \right) \\
&= \underbrace{\frac{1}{\sqrt{-g}} \partial_t
  \left( \sqrt{-g} \; \Gamma^t \right)}_{\dot{B} \Gamma^t}
  + \underbrace{\frac{1}{\sqrt{-g}} \partial_r
  \left( \sqrt{-g} \; \Gamma^r \right)}_{\left(A'+\frac{2}{r}\right) \Gamma^r}
  \nonumber\\
&\bleq + \underbrace{\frac{1}{\sqrt{-g}} \partial_\theta \left( \sqrt{-g}
  \; \Gamma^\theta \right)}_{\cot \theta \, \Gamma^t
  - \frac{\e^{B}}{r} \Gamma^r}
  + \underbrace{\frac{1}{\sqrt{-g}} \partial_\varphi
  \left( \sqrt{-g} \; \Gamma^\varphi \right)}_{-\frac{\gamma^2 \cos \varphi
  + \gamma^3 \sin \varphi}{r \sin \theta}} \nonumber\\
&= \dot{B} \Gamma^t + \left(A' + \frac{2}{r}\right) \Gamma^r
  - \frac{\e^B}{r} \Gamma^r \nonumber\\
&\bleq - \frac{1}{r \sin \theta} \left(\gamma^1 \cos \theta \sin \theta
  - (1 + \cos^2 \theta) (\gamma^2 \cos \varphi
  + \gamma^3 \sin \varphi) \right) \\
&= \dot{B} \Gamma^t
  + \left(A' + \frac{2}{r} \left(1 - \e^B\right) \right) \Gamma^r
\end{align}
and therefore
\begin{equation}
 \slashed{D} = \mathrm{i} \Gamma^t \left(\partial_t + \frac{\dot{B}}{2} \right)
  + \mathrm{i} \Gamma^r \left(\partial_r + \frac{A'}{2} + \frac{1 - \e^B}{r} \right)
  + \mathrm{i} \Gamma^\theta \partial_\theta + \mathrm{i} \Gamma^\varphi \partial_\varphi.
\end{equation}

Now we make use of the expansion \eqref{eqn:ansatz-fields} for $\psi$ as
before and insert
it into the Dirac equation \eqref{eqn:the-dirac-equation}. We introduce the
matrices $\vec{\widetilde{\alpha}}$ analogously to the previously defined 
$\vec{\alpha}$
\begin{subequations}%
\begin{align}
 \widetilde{\alpha}^r &= \gamma^0 \gamma^1 \cos \theta + \gamma^0 \gamma^2
  \sin \theta \cos \varphi + \gamma^0 \gamma^3 \sin \theta \sin \varphi \\
\widetilde{\alpha}^\theta &= -\gamma^0 \gamma^1 \sin \theta + \gamma^0 \gamma^2
  \cos \theta \cos \varphi + \gamma^0 \gamma^3 \cos \theta \sin \varphi \\
\widetilde{\alpha}^\varphi &= - \gamma^0 \gamma^2 \sin \varphi
  + \gamma^0 \gamma^3 \cos \varphi,
\end{align}%
\end{subequations}
as well as the rotated Pauli matrices
\begin{subequations}%
\label{eqn:sigma-tilde}%
\begin{align}
 \widetilde{\sigma}^r &= \sigma^1 \cos \theta
  + \sigma^2 \sin \theta \cos \varphi + \sigma^3 \sin \theta \sin \varphi \\
\widetilde{\sigma}^\theta &= -\sigma^1 \sin \theta
  + \sigma^2 \cos \theta \cos \varphi + \sigma^3 \cos \theta \sin \varphi \\
\widetilde{\sigma}^\varphi &= - \sigma^2 \sin \varphi + \sigma^3 \cos \varphi.
\end{align}%
\end{subequations}
Multiplying the Dirac equation \eqref{eqn:the-dirac-equation} with
$\gamma^0 = \beta$ from the left we then get
\begin{align}
 0 &= \exp\left(\frac{\mathrm{i} c^2}{\hbar} S\right) \frac{c^2}{\hbar}
  \sum_{n=0}^\infty \left(\frac{\sqrt{\hbar}}{c}\right)^n \Bigg[
  -\frac{1}{\sqrt{\hbar}} \e^{-A} \dot{S} a_{n-1}
  + \frac{\mathrm{i}}{\sqrt{\hbar}} \e^{-A} \dot{a}_{n-3} \nonumber\\
&\bleq + \frac{\mathrm{i}}{2 \sqrt{\hbar}} \e^{-A} \dot{B} a_{n-3}
  - \vec{\widetilde{\alpha}} \cdot (\grad{S}) a_n
  - \left(\e^{-B} - 1\right) \widetilde{\alpha}^r S' a_n
  + \mathrm{i} \vec{\widetilde{\alpha}} \cdot \grad a_{n-2} \nonumber\\
&\bleq + \mathrm{i} \left(\e^{-B} - 1\right) \widetilde{\alpha}^r a_{n-2}'
  + \frac{\mathrm{i}}{2} \e^{-B} A' \widetilde{\alpha}^r a_{n-2}
  + \frac{\mathrm{i}}{r} \left(\e^{-B} - 1\right) \widetilde{\alpha}^r a_{n-2} 
  \nonumber\\
&\bleq - \beta \frac{m}{\sqrt{\hbar}} a_{n-1} \Bigg]
\end{align}
and with the expansion for the exponentials \eqref{eqn:exponentials-expansion}
we have for each $n$
\begin{align}
 0 &= \sqrt{\hbar} \vec{\widetilde{\alpha}} \cdot (\grad{S}) a_n
+ \beta m a_{n-1}
- \mathrm{i} \sqrt{\hbar} \vec{\widetilde{\alpha}} \cdot \grad a_{n-2} \nonumber\\
&\bleq + \sum_{k=0}^n \Bigg[ C_k \dot{S} a_{n-k-1} 
- \mathrm{i} C_k \dot{a}_{n-k-3} 
- \frac{\mathrm{i}}{2} E_k a_{n-k-3} 
+ \sqrt{\hbar} D_k \widetilde{\alpha}^r S' a_{n-k} \nonumber\\
&\bleq - \mathrm{i} \sqrt{\hbar} D_k \widetilde{\alpha}^r a_{n-k-2}' 
- \frac{\mathrm{i} \sqrt{\hbar}}{2} F_k \widetilde{\alpha}^r a_{n-k-2} 
- \frac{\mathrm{i} \sqrt{\hbar}}{r} D_k \widetilde{\alpha}^r a_{n-k-2}
\Bigg] \\
\label{eqn:dirac-order-n}
&= \sqrt{\hbar} \vec{\widetilde{\alpha}} \cdot (\grad{S}) a_n
+ \beta m a_{n-1}
+ \dot{S} a_{n-1}
- \mathrm{i} \sqrt{\hbar} \vec{\widetilde{\alpha}} \cdot \grad a_{n-2}
- \mathrm{i} \dot{a}_{n-3} \nonumber\\
&\bleq + \sum_{k=1}^n \Bigg[ \sqrt{\hbar} D_k \widetilde{\alpha}^r S' a_{n-k}
+ C_k \dot{S} a_{n-k-1} 
- \mathrm{i} \sqrt{\hbar} D_k \widetilde{\alpha}^r a_{n-k-2}' \nonumber\\
&\bleq - \frac{\mathrm{i} \sqrt{\hbar}}{2}
\left(F_k + \frac{2}{r} D_k\right) \widetilde{\alpha}^r a_{n-k-2} 
- \mathrm{i} C_k \dot{a}_{n-k-3} 
- \frac{\mathrm{i}}{2} E_k a_{n-k-3} 
\Bigg]\,.
\end{align}

As before, at order $n=0$ this yields $\grad S = 0$, i.\,e. $S$ is a function
of time only, and at order $n=1$ we get $(\beta m + \dot{S}) a_0 = 0$.
Again, we split the four component spinors
$a_n = (a_{n,1}, a_{n,2}, a_{n,3}, a_{n,4})$
into two two component spinors
$a_n^>  = (a_{n,1}, a_{n,2})$ and $a_n^<  = (a_{n,3}, a_{n,4})$
and choose $a_0^< \equiv 0$. We are then left with $S = -mt$ and from
\eqref{eqn:dirac-order-n} we get the following two equations:
\begin{align}
\label{eqn:einstein-dirac-the-first-equation}
 0 &= \mathrm{i} \sqrt{\hbar} \vec{\widetilde{\sigma}} \cdot \grad a_{n-2}^<
  + \mathrm{i} \dot{a}_{n-3}^> + \sum_{k=1}^n \Big[ m C_k a_{n-k-1}^>
  + \mathrm{i} \sqrt{\hbar} D_k \widetilde{\sigma}^r {a_{n-k-2}^<}' \nonumber\\
 &\bleq + \frac{\mathrm{i} \sqrt{\hbar}}{2} \left( F_k + \frac{2}{r} D_k \right) 
  \widetilde{\sigma}^r a_{n-k-2}^< + \mathrm{i} C_k \dot{a}_{n-k-3}^>
  + \frac{\mathrm{i}}{2} E_k a_{n-k-3}^> \Big] \\
\label{eqn:einstein-dirac-the-second-equation}
a_{n-1}^< &= -\frac{\mathrm{i} \sqrt{\hbar}}{2 m} \vec{\widetilde{\sigma}}
  \cdot \grad a_{n-2}^> - \frac{\mathrm{i}}{2 m} \dot{a}_{n-3}^< - \sum_{k=1}^n
  \Big[ \frac{1}{2} C_k a_{n-k-1}^< + \frac{\mathrm{i} \sqrt{\hbar}}{2 m} D_k
  \widetilde{\sigma}^r {a_{n-k-2}^>}' \nonumber\\
&\bleq + \frac{\mathrm{i} \sqrt{\hbar}}{4 m} \left( \hspace{-1pt}F_k + \frac{2}{r} D_k \right)
  \widetilde{\sigma}^r a_{n-k-2}^> + \frac{\mathrm{i}}{2 m} C_k \dot{a}_{n-k-3}^<
  + \frac{\mathrm{i}}{4 m} E_k a_{n-k-3}^< \Big].
\end{align}

At order $n=2$ the first equation yields $0 = m C_1 a_0^>$ and therefore
$C_1 = 0$. This means $A_1 = 0$, $C_2 = - A_2$ and $F_1 = 0$. The second
equation yields
\begin{equation}
\label{eqn:a1_small_component}
 a_1^< = -\frac{\mathrm{i} \sqrt{\hbar}}{2 m} \vec{\widetilde{\sigma}}
  \cdot \grad a_0^>.
\end{equation}

Inserting this into the first equation at order $n=3$ we get
\begin{equation}
 0 = \mathrm{i} \sqrt{\hbar} \vec{\widetilde{\sigma}} \cdot \grad \left( 
  -\frac{\mathrm{i} \sqrt{\hbar}}{2 m} \vec{\widetilde{\sigma}} 
  \cdot \grad a_0^> \right) + \mathrm{i} \dot{a}_0^> - m A_1 a_0^>
\end{equation}
As $(\vec{\widetilde{\sigma}} \cdot \vec{u})^2 = \vec{u}^2$ for any vector
$\vec{u}$ we obtain the \sne
\begin{equation}
 \mathrm{i} \hbar \dot{a}_0^> = -\frac{\hbar^2}{2 m} \Delta a_0^> + V a_0^>,
\end{equation}
with the potential $V = m \hbar A_2$ as we did for the Klein-Gordon equation.

The second equation at order $n=3$ yields
\begin{equation}
 a_2^< = -\frac{\mathrm{i} \sqrt{\hbar}}{2 m} \vec{\widetilde{\sigma}}
  \cdot \grad a_1^>
  + \frac{\mathrm{i} \sqrt{\hbar}}{2 m} B_1 \widetilde{\sigma}^r \left( {a_0^>}'
  + \frac{1}{r} a_0^> \right).
\end{equation}
This result could now again be inserted into the first equation
\eqref{eqn:einstein-dirac-the-first-equation} at order $n=4$
which would result in the evolution equation for $a_1^>$, but
in contrast to the pure Dirac equation (section \ref{sec:dirac-nr})
where we obtained the
Pauli equation also for $a_1^>$, this evolution equation
will be different from the \sne.

\subsubsection{Einstein's equations}
Let us first derive the stress-energy-tensor for the Dirac field.
The Dirac-Lagrangian is
\begin{equation}
 \mathcal{L} = \hbar c \overline{\psi} \left( \slashed{D}
  - \frac{m c}{\hbar} \right) \psi \sqrt{-g}.
\end{equation}
Considering the variation with respect to $g^{\mu \nu}$ on-shell
(i.\,e. assuming the Dirac equation to be satisfied) we have
\begin{equation}
 \delta \mathcal{L} = \hbar c \;\mathrm{Re} \left[ \overline{\psi}
  \left(\mathrm{i} \left(\delta \Gamma^\mu\right) \partial_\mu
  + \delta Y\right) \psi \sqrt{-g} \right].
\end{equation}
According to Finster \emph{et al.}~\cite{Finster:1999}, with our special
choice \eqref{eqn:choice-dirac-matrices} for the Dirac matrices
\begin{align}
 \mathrm{Re} \left(\overline{\psi} \delta Y \psi \right)
  &= \frac{1}{16} \epsilon^{\mu \nu \rho \sigma}
  \left(\delta g_{\nu \lambda}\right)
  \tr{\Gamma^\lambda \partial_\mu \Gamma_\rho} \overline{\psi}
  \gamma^5 \Gamma_\sigma \psi \\
\label{eqn:delta_Gamma}
 \delta \Gamma^\mu &= -\frac{1}{2} g^{\mu \nu}
  \left(\delta g_{\nu \lambda}\right) \Gamma^\lambda \\
 \delta g_{\mu \nu} &= - g_{\mu \lambda} g_{\nu \sigma}
  \delta g^{\lambda \sigma}
\end{align}
and therefore the stress-energy-tensor is
\begin{multline}
 T_{\mu \nu} = -\frac{\hbar c}{2} \;\mathrm{Re} \left[\overline{\psi}
  \left(\mathrm{i} \Gamma_\nu \partial_\mu
  + \mathrm{i} \Gamma_\mu \partial_\nu \right) \psi \right] \\
+ \frac{\hbar c}{8} \epsilon^{\alpha \lambda \rho \beta}
  \left[g_{\mu \lambda} \tr{\Gamma_\nu \partial_\alpha \Gamma_\rho}
  + g_{\nu \lambda} \tr{\Gamma_\mu \partial_\alpha \Gamma_\rho} \right] 
  \overline{\psi} \gamma^5 \Gamma_\beta \psi.
\end{multline}
Note that \eqref{eqn:delta_Gamma} is determined up to local Lorentz
transformation. $\delta \Gamma^\mu$ has to obey
\begin{equation}
 \delta \akomm{\Gamma^\mu}{\Gamma^\nu}
  = - \delta g^{\mu \nu} \quad
  \Rightarrow \quad \akomm{\delta \Gamma^\mu}{\Gamma^\nu}
  = - \delta g^{\mu \nu}
\end{equation}
and \eqref{eqn:delta_Gamma} is a special solution to this equation.

The non-vanishing components of the stress-energy-tensor for a Dirac
field $\psi$ are then given by
\begin{subequations}%
\begin{align}
T_{tt}        &= -\hbar c \mathrm{Re} \left[-\mathrm{i} \e^{2A} c^2 \overline{\psi}
                 \Gamma^t \dot{\psi}\right] \\
T_{rr}        &= -\hbar c \mathrm{Re} \left[\mathrm{i} \e^{2B} \overline{\psi}
                 \Gamma^r \psi'\right] \\
T_{tr}        &= -\frac{\hbar c}{2} \mathrm{Re} \left[-\mathrm{i} \e^{2A} c^2
                 \overline{\psi} \Gamma^t \psi'
                 + \mathrm{i} \e^{2B} \overline{\psi} \Gamma^r \dot{\psi}\right] \\
T_{t \theta}  &= -\frac{\hbar c}{2} \mathrm{Re} \left[\mathrm{i} r^2 \overline{\psi}
                 \Gamma^\theta \dot{\psi}
                 - \mathrm{i} \e^{2A} c^2 \overline{\psi} \Gamma^t
                 \partial_\theta \psi \right] \nonumber \\
                 &\bleq -\frac{\hbar c^2}{2} \e^{A-B} r^2 \sin \theta
                 \left(A' - \frac{1-\e^B}{r}\right) \overline{\psi}
                 \gamma^5 \Gamma^\varphi \psi \\
T_{t \varphi} &= -\frac{\hbar c}{2} \mathrm{Re} \left[\mathrm{i} r^2 \sin^2 \theta 
                 \overline{\psi} \Gamma^\varphi \dot{\psi}
                 - \mathrm{i} \e^{2A} c^2 \overline{\psi} \Gamma^t
                 \partial_\varphi \psi \right] \nonumber \\
                 &\bleq + \frac{\hbar c^2}{2} \e^{A-B} r^2 \sin \theta 
                 \left(A' - \frac{1-\e^B}{r}\right) \overline{\psi}
                 \gamma^5 \Gamma^\theta \psi \\
T_{r \theta}  &= -\frac{\hbar c}{2} \mathrm{Re} \left[\mathrm{i} r^2 \overline{\psi}
                 \Gamma^\theta \psi'
                 + \mathrm{i} \e^{2B} \overline{\psi} \Gamma^r \partial_\theta \psi \right] -\frac{\hbar}{2} \e^{B-A} r^2
                 \sin \theta \dot{B} \overline{\psi} \gamma^5
                 \Gamma^\varphi \psi \\
T_{r \varphi} &= -\frac{\hbar c}{2} \mathrm{Re} \left[\mathrm{i} r^2 \sin^2 \theta 
                 \overline{\psi} \Gamma^\varphi \psi'
                 + \mathrm{i} \e^{2B} \overline{\psi} \Gamma^r \partial_\varphi \psi \right]
                 + \frac{\hbar}{2} \e^{B-A} r^2 \sin \theta \dot{B} 
                 \overline{\psi} \gamma^5 \Gamma^\theta \psi \\
T_{\theta \theta} &= -\hbar c \mathrm{Re} \left[ \mathrm{i} r^2 \overline{\psi} \Gamma^\theta \partial_\theta \psi \right] \\
T_{\theta \varphi} &= -\frac{\hbar c}{2} \mathrm{Re} \left[ \mathrm{i} r^2 \overline{\psi} \Gamma^\theta \partial_\varphi \psi + \mathrm{i} r^2 \sin^2 \theta \overline{\psi} \Gamma^\varphi \partial_\theta \psi \right] \\
T_{\varphi \varphi} &= -\hbar c \mathrm{Re} \left[ \mathrm{i} r^2 \sin^2 \theta \overline{\psi} \Gamma^\varphi \partial_\varphi \psi \right].
\end{align}%
\end{subequations}

Up to order $c^0$ these can be simplified to
\begin{subequations}%
\begin{align}
T_{tt}        &= -\hbar c \mathrm{Re} \left[\psi^\dagger
                 \left(-\mathrm{i} \e^{A} c \right) \dot{\psi}\right] \\
T_{rr}        &= -\hbar c \mathrm{Re} \left[\psi^\dagger
                 \left(\mathrm{i} \e^{B} \widetilde{\alpha}^r \right) \psi'\right] \\
T_{tr}        &= -\frac{\hbar c}{2} \mathrm{Re} \left[\psi^\dagger
                 \left(-\mathrm{i} \e^{A} c \right) \psi'
                 + \psi^\dagger \left(\mathrm{i} \e^{B} \widetilde{\alpha}^r \right)
                 \dot{\psi}\right] \\
T_{t \theta}  &= -\frac{\hbar c}{2} \mathrm{Re} \left[\psi^\dagger
                 \left(\mathrm{i} r \widetilde{\alpha}^\theta \right) \dot{\psi}\right]
                 + \order{c^0} \\
T_{r \theta}  &= -\frac{\hbar c}{2} \mathrm{Re} \left[\psi^\dagger
                 \left(\mathrm{i} r \widetilde{\alpha}^\theta \right) \psi'
                 + \psi^\dagger \left(\mathrm{i} \e^{B} \widetilde{\alpha}^r \right) \partial_\theta \psi \right] + \order{c^0} \\
T_{t \varphi}  &= -\frac{\hbar c}{2} \mathrm{Re} \left[\psi^\dagger
                 \left(\mathrm{i} r \sin \theta \widetilde{\alpha}^\varphi \right) \dot{\psi}\right]
                 + \order{c^0} \\
T_{r \varphi}  &= -\frac{\hbar c}{2} \mathrm{Re} \left[\psi^\dagger
                 \left(\mathrm{i} r \sin \theta \widetilde{\alpha}^\varphi \right) \psi'
                 + \psi^\dagger \left(\mathrm{i} \e^{B} \widetilde{\alpha}^r \right) \partial_\varphi \psi \right] + \order{c^0} \\
T_{\theta \theta}  &= -\hbar c \mathrm{Re} \left[\psi^\dagger
                 \left(\mathrm{i} r \widetilde{\alpha}^\theta \right) \partial_\theta \psi \right] \\
T_{\theta \varphi}  &= -\frac{\hbar c}{2} \mathrm{Re} \left[\psi^\dagger
                 \left(\mathrm{i} r \widetilde{\alpha}^\theta \right) \partial_\varphi \psi + \psi^\dagger
                 \left(\mathrm{i} r \sin \theta \widetilde{\alpha}^\varphi \right) \partial_\theta \psi \right] \\
T_{\varphi \varphi}  &= -\hbar c \mathrm{Re} \left[\psi^\dagger
                 \left(\mathrm{i} r \sin \theta \widetilde{\alpha}^\varphi \right) \partial_\varphi \psi \right].
\end{align}%
\end{subequations}

Again, as for the Klein-Gordon stress-energy tensor, we expand
$T_{\mu \nu} / c^4$ using \emph{Mathematica}
and consider Einstein's
equations with the Einstein tensor given in equation
\eqref{eqn:Einstein-tensor} for each component order by order, making use of
$A_1 = 0$. The result differs from the Klein-Gordon result only slightly.

\paragraph{\boldmath$n=0$:}
The $tt$-component yields $(r B_1)' = 0$.

\paragraph{\boldmath$n=1$:}
The $tt$-component yields
\begin{equation}
\label{eqn:Dirac-n1-tt}
 \frac{2 \hbar}{r^2} \left(\frac{3}{2} B_1^2 + (r B_2)' \right)
    = 8 \pi G m \abs{a_0^>}^2.
\end{equation}

\paragraph{\boldmath$n=2$:}
The $tt$-component yields
\begin{equation}
\label{eqn:Dirac-n2-tt}
 \frac{2 \hbar}{r^2} \left(-4 B_1^3 - 3 r B_1 B_2' + (r B_3)' \right)
  = 8 \pi G m \left({a_1^>}^\dagger a_0^> + {a_0^>}^\dagger a_1^> \right).
\end{equation}
The $rr$-components yields $B_1=0$, the $tr$-component $\dot{B}_1 = 0$ and the
$\theta \theta$-component $B_1'=0$ are then trivial as well as the order $n=0$
equation. Equations \eqref{eqn:Dirac-n1-tt} and \eqref{eqn:Dirac-n2-tt} then 
simplify to
\begin{align}
(r B_2)' &= \frac{4 \pi G m r^2}{\hbar} \abs{a_0^>}^2 \\
\label{eqn:Dirac-n2-tt-simplified}
(r B_3)' &= \frac{4 \pi G m r^2}{\hbar} \left({a_1^>}^\dagger a_0^> + {a_0^>}^\dagger a_1^> \right)
\end{align}

\paragraph{\boldmath$n=3$:}
The $tt$-component yields
\begin{multline}
\label{eqn:Dirac-n3-tt}
4 (A_2-B_2) (r B_2)' + B_2^2 - 2 r B_2 B_2' + (r B_4)'
  = \frac{4 \pi G r^2}{\hbar^2} \Big[ \frac{\mathrm{i} \hbar}{2}
  \left( {a_0^>}^\dagger \dot{a}_0^> - \dot{a}_0^>{}^\dagger a_0^> \right) \\
+ m \hbar \left( A_2 \abs{a_0^>}^2 + \abs{a_1}^2 + {a_2^>}^\dagger a_0^>
  + {a_0^>}^\dagger a_2^> \right) \Big].
\end{multline}
The $rr$-component yields
\begin{equation}
 B_2 = r A_2'
\end{equation}
and the $\theta \theta$-component is just the derivative of the $rr$-component.
The $tr$-component yields (with $\widetilde{\sigma}^r$ as defined in
\eqref{eqn:sigma-tilde})
\begin{align}
\label{eqn:dirac-final-dotB}
\frac{2}{r} \dot{B}_2 &= 2 \pi \mathrm{i} G \left({a_0^>}^\dagger {a_0^>}'
			  - {a_0^>}'{}^\dagger a_0^> \right)
			  - \frac{8 \pi G m}{\sqrt{\hbar}} \mathrm{Re}
			  \left(a_0^>{}^\dagger \widetilde{\sigma}^r a_1^< \right)
\nonumber\\
&= 2 \pi \mathrm{i} G \left({a_0^>}^\dagger {a_0^>}' - {a_0^>}'{}^\dagger a_0^> \right)
   + 4 \pi G \mathrm{Re} \left[a_0^>{}^\dagger
     \left( \mathrm{i} \grad + \vec{\sigma} \times \grad \right)_r a_0^> \right],
\nonumber\\
&= 4 \pi \mathrm{i} G \left({a_0^>}^\dagger {a_0^>}' - {a_0^>}'{}^\dagger a_0^> \right)
   + 4 \pi G \mathrm{Re} \left[a_0^>{}^\dagger
     \left( \vec{\sigma} \times \grad \right)_r a_0^> \right],
\end{align}
where we made use of \eqref{eqn:a1_small_component} to express
$a_1^<$ in terms of $a_0^>$ in the second line.
The $t \theta$- and $t \varphi$-components yield
\begin{subequations}%
\label{eqn:t-theta-t-phi}%
\begin{alignat}{2}
\label{eqn:t-theta-t-phi-a}
 0 &= \mathrm{Re} \left(a_0^>{}^\dagger \widetilde{\sigma}^\theta a_1^< \right) &
 \quad \Leftrightarrow \quad
 0 &= \mathrm{Re} \big[a_0^>{}^\dagger
     \left( \mathrm{i} \grad + \vec{\sigma} \times \grad \right)_\theta a_0^> \big] \\
 0 &= \mathrm{Re} \left(a_0^>{}^\dagger \widetilde{\sigma}^\varphi a_1^< \right)&
 \quad \Leftrightarrow \quad
 0 &= \mathrm{Re} \big[a_0^>{}^\dagger
     \left( \mathrm{i} \grad + \vec{\sigma} \times \grad \right)_\varphi a_0^> \big]\,.
\end{alignat}%
\end{subequations}

Now, because we consider the spherically symmetric Einstein's equations
we have to average over all spin orientations in order to get a symmetric
stress-energy-tensor. This implies that terms proportional to
$\vec{\sigma} \times \grad$ in equations \eqref{eqn:dirac-final-dotB}
and \eqref{eqn:t-theta-t-phi} will vanish and that \eqref{eqn:t-theta-t-phi}
is equivalent to vanishing $\theta$- and $\varphi$-components
of the probability current
\begin{equation}
 \vec{j}_\text{Dirac} = \frac{\mathrm{i} \hbar}{2 m}
			\left( \left({\grad a_0^>}{}^\dagger\right) a_0^>
			- {a_0^>}^\dagger \grad a_0^> \right)\,.
\end{equation}

If then, again, we define the potentials
$V = m \hbar A_2$ and $U = m \hbar B_2$ and define $j$ as the $r$-component of
the probability current
we get the same set of equations as in \eqref{eqn:set-pre-poisson}
\begin{align}
U &= r V' \\
(r U)' &= 4 \pi G m^2 r^2 \abs{a_0^>}^2 \\
\dot{U} &= -4 \pi G m^2 r j,
\end{align}
together with equations \eqref{eqn:Dirac-n2-tt-simplified} and
\eqref{eqn:Dirac-n3-tt}.
As shown in subsection \ref{sec:Poisson} this yields the Poisson equation
for the potential $V$.

\section{Summary}
From our analysis of the Klein-Gordon and Dirac equations 
we conclude that the expansion in either $\hbar$ or $1/c$ 
according to our ansatz \eqref{eqn:ansatz-fields} for the fields 
is a valid scheme to obtain both semi-classical and non-relativistic 
limits of field equations in an unambiguous way. Applying the 
same ansatz to the self-gravitating Klein-Gordon and Dirac fields, 
as mathematically represented by the Einstein-Klein-Gordon and  
Einstein-Dirac systems, leads to the \sne. Hence we may say that 
the \sne follows from the self-gravitating fields in the same 
way as the linear Schr\"odinger equation can be derived in 
flat space. Seen from that direction one concludes that the 
\sne should provide a better description than the linear 
equation. 

However, one may ask: description of what? In order to arrive 
at our result we considered the \emph{classical} (i.\,e. not 
second quantised) Klein-Gordon or Dirac field as source for 
the classical gravitational field. The Schr\"odinger function 
then merely appears as part of these fields in the $1/c$ 
expansion and with certain phases (due to the rest mass) 
subtracted. The central question is, whether this is the 
right way to represent the gravitational field of quantum 
systems. Reading the classical fields as one-particle 
amplitudes, it amounts to assuming the validity of the 
semi-classical Einstein equations 
\begin{equation}
\label{eq:SemiclEinsteinEq}  
R_{\mu\nu}-\tfrac{1}{2}R\,g_{\mu\nu}=\frac{8\pi G}{c^4}
\langle\Psi\vert\hat T_{\mu\nu}\vert\Psi\rangle
\end{equation}
where $\Psi$ now represents the full (second quantised) field. 
This equation has a long and controversial history, which we will 
not review here. The interested reader is referred to the book 
by Kiefer \cite{Kiefer:QuantumGravity} and references therein,
and the recent discussion in \citet{Carlip:2008,Salzman.Carlip:2006}.
Interesting early (1957) discussions between Feynman and others 
on that subject may be found in \cite{Conference.NorthCarolina:1957}.

An obvious objections is this: If there were a 
gravitational self-coupling of the Schr\"odinger function, then 
there should also be an electromagnetic self-coupling. 
In a lowest $1/c$ expansion (neglecting magnetic fields) this 
would again result in equation~\eqref{eq:SchroedingerNewton} 
with $G m^2$ replaced by $-e^2/(4\pi\epsilon_0)$, leading to enhanced 
dispersion due to electrostatic repulsion. Applied to the 
hydrogen atom, where the electron now not only ``sees'' the 
electrostatic attraction of the proton but also the electrostatic 
repulsion of its own charge-cloud, it seems obvious that this 
cannot again reproduce the known energy spectrum. Interestingly, 
precisely this idea of implementing the electromagnetic 
self-interaction of the quantum-mechanical wave function 
occurred to Schr\"odinger immediately after he wrote his 
famous papers on wave mechanics. In 1927 he argued  
\cite{Schroedinger:1927} that, as a matter of principle, 
the self-coupling was required by consistency in order to 
get closed systems of field equations. Just like one derives 
the radiation reaction of charges through interaction with 
their own field in ordinary Maxwell theory. But being convinced 
that the result of this was incompatible with observed facts, 
like the hydrogen energy levels, Schr\"odinger concludes that, 
from a classical field-theoretic point of view, there is a 
``strange violation of the closedness of the field equations''%
 \footnote{``Gerade die \emph{Geschlossenheit} der Feldgleichungen 
erscheint somit in eigenartiger Weise durchbrochen'' (p.\,271 \cite{Schroedinger:1927}).}.

Schr\"odinger's idea has been revived in the mid 1980's by Barut 
and collaborators, who wrote down a rather obvious non-linear 
Dirac equation which is obtained from the linear equation by
first splitting the field into a self- and external part and 
then eliminating the self-part by means of Green's functions and 
the self-current, the latter then introducing non-linearities. 
They claimed that without any further input from QED this  
suffices to account for effects usually attributed to the 
quantisation of the electromagnetic field, like Lamb shift, 
spontaneous emission, and  anomalous $g-2$. And even though 
from a path-integral perspective their equation is, in fact, 
just that one obtains from an effective action after 
integrating out the photons, it seems very far reaching indeed.
What is more, it was also suggested (but not shown) that 
the correct values of the hydrogen spectrum can be obtained
if due care is taken of the non-perturbative nature of this 
self interaction~\cite{Barut:1988}. 

Another problem concerns the application of the \sne to 
molecular interferometry. What aspect of the complex 
molecule, comprising many degrees of freedom, does our 
$\psi$ represent? It just depends on 3 coordinates, so one 
might be tempted to identify it with the centre-of-mass 
motion. But one would certainly not expect an influence of 
the centre-of-mass motion from gravitational pair-interactions
of many-particle systems~\cite{Adler:2007}. Self-interactions 
are also absent in a manifest Galilei invariant (i.e. 
manifest ``non-relativistic'') second-quantised theory 
of the Schr\"odinger field in a Newton-Cartan spacetime~\cite{Christian:1997}. This still leaves open the possibility to think of the \sne as 
fundamental deviation from one-particle quantum mechanics.
Alternatively one can regard our $\psi$ as a condensate and 
think of the non-linearity as coming about through effective 
interactions, just like in the Choquard equation~\cite{Lieb:1977,Froehlich:2003}.

Our results concerning the derivability of the \sne 
by WKB-like methods from the Einstein-Klein-Gordon and 
Einstein-Dirac systems are, as such, independent of these questions.
But, as stressed in the introduction, they clearly need to 
be addressed, in one form or another, in any attempt to 
make a well founded physical applications. 

\section*{Acknowledgements}
We acknowledge funding through the Centre for Quantum Engineering
and Space-Time Research (\small{QUEST}) at the Leibniz University Hannover.


\end{document}